\documentclass[prd,twocolumn,amsmath,amssymb,floatfix,superscriptaddress,nofootinbib]{revtex4-1}

\usepackage{graphicx}
\usepackage{scrextend}
\usepackage{amssymb}
\usepackage{amsmath}
\usepackage{bm}
\usepackage{color}
\usepackage{multirow}
\usepackage{cprotect}

\DeclareMathAlphabet\mathbfcal{OMS}{cmsy}{b}{n}
\definecolor{darkgreen}{RGB}{50,150,0}
\definecolor{purple}{cmyk}{0.5,1.0,0,0}

\def\edth{\;\raise1.0pt\hbox{$'$}\hskip-6pt\partial}
\def\baredth{\;\overline{\raise1.0pt\hbox{$'$}\hskip-6pt
\partial}}

\def\be{\begin{equation}}
\def\ee{\end{equation}}
\def\ben{\begin{equation} \nonumber}
\def\een{\end{equation}}
\def\ban{\begin{eqnarray*}}
\def\ean{\end{eqnarray*}}
\def\ba{\begin{eqnarray}}
\def\ea{\end{eqnarray}}
\def\({\left(}
\def\){\right)}

\newcommand{\PP}{{\phi\phi}}

\definecolor{ultramarine}{rgb}{0.07, 0.04, 0.56}
\definecolor{cadmiumgreen}{rgb}{0.0, 0.42, 0.24}
\definecolor{indigo(dye)}{rgb}{0.0, 0.25, 0.42}
\usepackage[linktocpage=true]{hyperref}
\hypersetup{
colorlinks=true,
citecolor=ultramarine,
linkcolor=cadmiumgreen,
urlcolor=indigo(dye),
pdfauthor={},
pdftitle={},
pdfsubject={}
}

\begin{document}

\title{Tensions between direct measurements of the lens power spectrum from Planck data}

\author{Pavel Motloch}
\affiliation{Kavli Institute for Cosmological Physics, Department of Physics, University of Chicago, Chicago, Illinois 60637, U.S.A}

\author{Wayne Hu}
\affiliation{Kavli Institute for Cosmological Physics, Department of Astronomy \& Astrophysics,  Enrico Fermi Institute, University of Chicago, Chicago, Illinois 60637, U.S.A}

\begin{abstract}
\noindent
We apply a recently developed method to directly measure the
gravitational lensing power spectrum from CMB power spectra
 to the Planck satellite data.   This method allows us to analyze the tension
between the temperature power spectrum and lens reconstruction in a
model independent way.  
Even when allowing for arbitrary variations in the lensing power spectrum,
the
tension remains at the  2.4$\sigma$ level.  
By separating the lensing and unlensed high redshift information in the
CMB power spectra, 
we also show that under $\Lambda$CDM the two are in tension at a 
similar level whereas the unlensed information is consistent with lensing
reconstruction.   These anomalies are driven by the smoother
acoustic peaks relative to $\Lambda$CDM at $\ell \sim 1250 - 1500$.
Both tensions relax slightly when polarization data are considered. 
This technique also isolates the one combination of the lensing power
spectrum multipoles that the Planck CMB power spectra currently constrain and
can be straightforwardly generalized to future data when CMB 
power spectra constrain multiple aspects of lensing which are themselves
correlated with lensing reconstruction.

\end{abstract}

\maketitle

\section{Introduction}
\label{sec:intro}

Measurements of anisotropies in the cosmic microwave background (CMB) have helped to establish $\Lambda$CDM
as the standard cosmological model and measure its
parameters with high precision. Currently, CMB data are precise enough to detect the
effects of gravitational lensing (see \cite{Lewis:2006fu} for a review) at high
significance \cite{Smith:2007rg, Hirata08,Hanson:2013hsb, Das:2011ak, Keisler:2011aw,
Planck2013XVII, Keisler:2015hfa,Ade:2015zua, Array:2016afx, Sherwin2016}.
This secondary signal depends on growth of structure in the
universe, which can be leveraged to  measure better certain parameters in $\Lambda$CDM like the sum of the neutrino masses
 and search for
new physics beyond {the standard cosmological model}.

Information carried by the lensing potential $\phi$ can be recovered either by measuring its effect
on CMB power spectra, in particular the smoothing of the acoustic peaks
\cite{Seljak:1995ve}, or by higher point combinations  of the temperature and polarization maps.  The latter is possible, because
gravitational lensing generates a correlation between measured CMB fields and their
gradients  \cite{Zaldarriaga:2000ud}.  This correlation
can be used to reconstruct $\phi$, for
example using quadratic combinations of maps \cite{Hu:2001fa,Hu:2001kj,Okamoto:2003zw} or iterative approaches
\cite{Hirata:2003ka,Smith:2010gu}. The reconstructed potential then serves as a new
 observable whose power spectrum $C_L^\PP$ contains cosmological information.

Within the context of $\Lambda$CDM, measurements of the Planck satellite show internal
tension -- at significance over 2$\sigma$ -- in the amount of lensing apparently present in
the temperature power spectra data \cite{Ade:2015xua, Aghanim:2016sns,Addison:2015wyg,Ade:2013zuv}. At
the same time, the amount of lensing detected through the lensing reconstruction seems to
be consistent with the $\Lambda$CDM model. Such tensions are interesting, as they could indicate  residual systematics in the data, which would have to be understood before
performing delensing using the reconstructed $\phi$ \cite{Carron:2017vfg}.
They might even 
represent hints of new physics beyond the $\Lambda$CDM model.

In this paper,  we perform a model-independent investigation of $C_L^\PP$ information in the Planck 
temperature and polarization power spectra and assess its tension with lensing reconstruction.
As opposed to previous studies of the tension, which considered changes to only the
amplitude of the gravitational lensing potential relative to the $\Lambda$CDM model, we allow for more general
changes. This enables us to probe the possibility that models beyond
$\Lambda$CDM might resolve this lensing tension.

Specifically, we implement the method introduced in 
\cite{Motloch:2016zsl, Motloch:2017rlk}, in which the lensing potential is constrained directly.
This method isolates  the precise aspect of $C_L^\PP$ that temperature and polarization power
spectra constrain, which enables a more incisive and model-independent assessment of any tension with reconstruction.
Additionally, this work represents the first test of the method on real data. 
This method will become increasingly 
useful in the future as CMB temperature and polarization power spectra constrain 
 more than just the amplitude of the gravitational lensing potential and enable  consistency tests that are largely 
 immune to cosmic variance.

This paper is organized as follows. In \S\ref{sec:analysis_details} we summarize the
Planck likelihoods used in this work and introduce our technique for probing the
gravitational lensing potential. In
\S\ref{sec:model_indept_lensing_constraints} we use this method to derive Planck lensing constraints 
from CMB power spectra and lens reconstruction.  
In \S\ref{sec:significance} we then
evaluate the significance of the tensions between the two and with $\Lambda$CDM.
We discuss our findings in \S\ref{sec:discussion}.

\section{Lensing methodology}
\label{sec:analysis_details}

In this section we first introduce the data and analysis method used in
this work. After that we detail the technique we use to directly measure the
gravitational lensing potential from the data; this technique is employed to probe Planck
lensing tensions in the later sections. 

\begin{table*}
\caption{Planck likelihoods used in this work} 
\label{tab:planck_likes}
\begin{tabular}{ccccc}
\hline\hline
Label & Power spectra & $\ell$-range & Binned? & Name\\
\hline
\verb|TT| & TT & $\ell \ge 30$ & yes &\verb|plik_dx11dr2_HM_v18_TT|\\
\verb|TTTEEE| & TT,TE,EE & $\ell \ge 30$ & yes & \verb|plik_dx11dr2_HM_v18_TTTEEE|\\
\verb|lowT| & TT & $\ell < 30$  & no &\verb|commander_rc2_v1.1_l2_29_B|\\
\verb|lowTEB| & TT,TE,EE,BB & $\ell < 30$ & no &\verb|lowl_SMW_70_dx11d_2014_10_03_v5c_Ap|\\
\verb|PP| & $\PP$ & $40 \le \ell \le 400$& yes & \verb|smica_g30_ftl_full_pp|\\
\verb|liteTT| & TT & $\ell \ge 30$ & yes & \verb|plik_lite_v18_TT|\\
\hline\hline
\end{tabular}
\end{table*}

\subsection{Data sets and MCMC}

For the analyses in this work we use the publicly released Planck 2015
likelihoods\footnote{https://www.cosmos.esa.int/web/planck/pla} for the power spectra of the CMB temperature, polarization and 
of the gravitational lensing potential reconstructed from their maps as summarized in 
Table~\ref{tab:planck_likes}. Joint use of multiple likelihoods will be denoted by a plus sign
connecting them.
For the  analyses using only the lensing reconstruction likelihood \verb|PP|, correction for the $N^{(0)}, N^{(1)}$ biases is performed
using the best fit power spectra to \verb|TT+lowTEB| likelihoods assuming the six parameter
$\Lambda$CDM model.  The likelihood \verb|liteTT|, which we use for the lens principal component
construction and checking robustness of the results below, is a high-$\ell$ temperature likelihood, with
the foreground parameters pre-marginalized over.    Otherwise, we marginalize over the standard foreground parameters with their default priors, where
applicable.

For the analyses in this paper we use the Markov Chain Monte Carlo (MCMC) code
CosmoMC\footnote{https://github.com/cmbant/CosmoMC} \cite{Lewis:2002ah} to sample the
posterior probability in the various parameter spaces described in the next section.
Each of our chains has a sufficient number of samples such that the Gelman-Rubin statistic $R-1$ \cite{Gelman:1992zz} falls below
0.01.

\subsection{Lens PC measurement technique}
\label{sec:pc_intro}

We can measure gravitational lensing of the CMB from both temperature and polarization
power spectra and from lensing reconstruction.
While lensing reconstruction maps directly measure the lens power spectrum $C_L^\PP$, in the standard analysis of
the temperature and polarization power lensing is inferred from the set
of model parameters.   This makes it difficult to compare these two distinct sources of lensing information 
directly since information from the latter is embedded in the cosmological parameter constraints.
 In this work we instead consider model-independent constraints on
$C_L^{\phi\phi}$. For this purpose we introduce a basis of
$N$ effective parameters $\Theta^{(i)}$ which determine arbitrary variations 
around a fixed fiducial power spectrum  $C_{L, \mathrm{fid}}^\PP$  as 
\be
\label{definition}
	C_L^\PP = C_{L, \mathrm{fid}}^\PP \exp\(\sum_{i = 1}^N K^{(i)}_L\, \Theta^{(i)} \) .
\ee
In this setup, constraining
$\Theta^{(i)}$ from the data  corresponds directly to constraining the gravitational lensing potential.
This should be contrasted with the common approach of introducing a phenomenological parameter $A_L$
which multiplies $C_L^{\phi\phi}$ at each point in the model  space and cannot be so interpreted once model parameters are marginalized over.   

\begin{figure}[b]
\center
\includegraphics[width = 0.49 \textwidth]{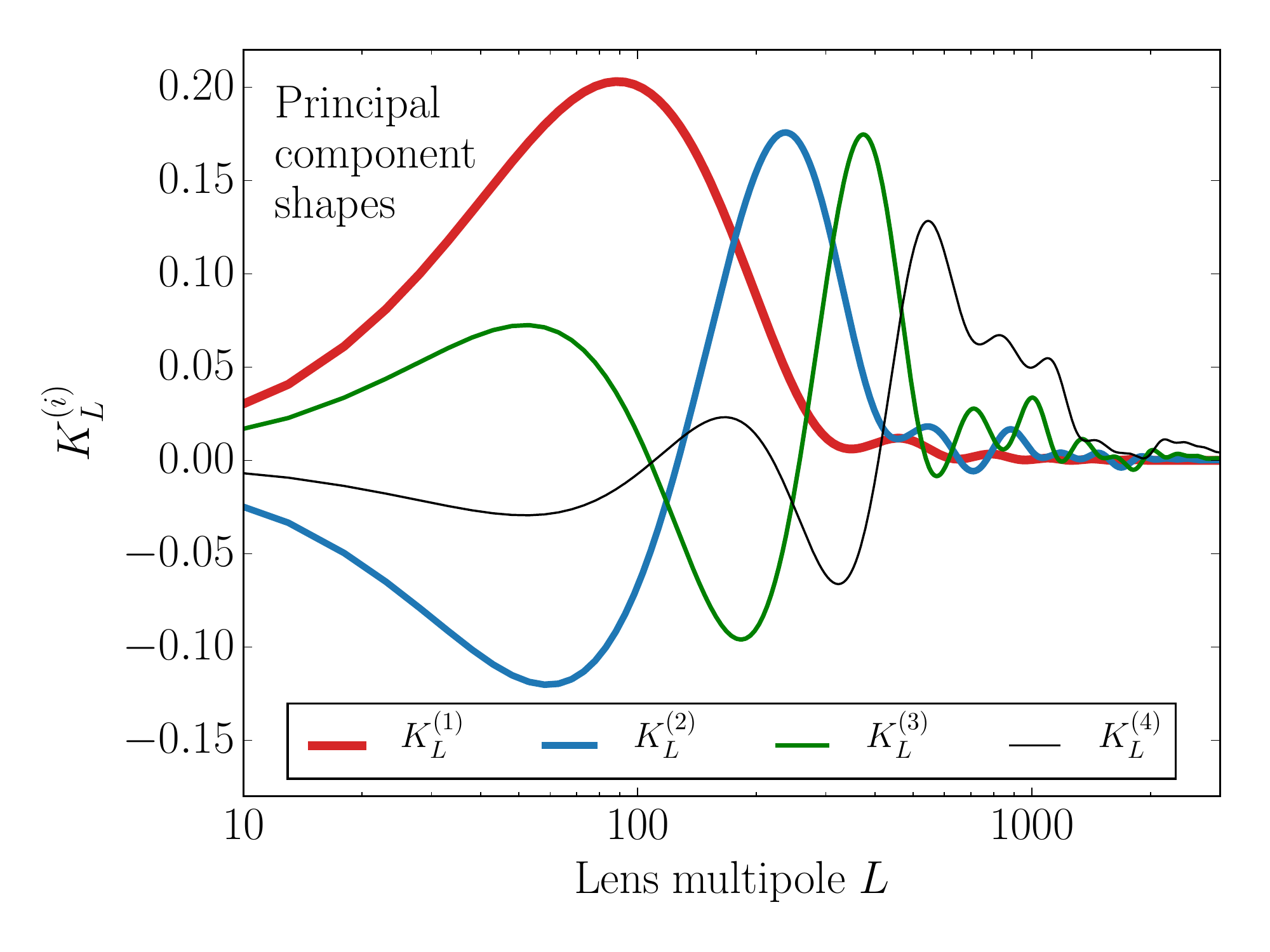}
\cprotect\caption{Functions $K_L^{(i)}$ corresponding to the four principal components
of the lens potential best measured by the Planck lensed TT power spectrum,
determined from the \verb|liteTT| likelihood. 
}
\label{fig:pcs} 
\end{figure}

We choose $K_L^{(i)}$ such that $\Theta^{(i)}$ correspond to $N$ principal components
(PCs) of
the gravitational lensing potential best measured by the Planck lensed TT power
spectrum. We determine them from the data covariance matrix provided with the
Planck likelihood \verb|liteTT| using a Fisher matrix construction; the resultant eigenmodes $K_L^{(i)}$ are shown in
Fig.~\ref{fig:pcs}. The lowest variance mode peaks around $L =100$ and is
much better constrained than the second best constrained component; the Fisher matrix
forecast predicts a factor of $\sim 50$ in the ratio of variances.  This is largely because
Planck data do not strongly constrain the $BB$ power spectrum \cite{Smith:2006nk}
or temperature multipoles above $\ell \approx 2000$, which are sensitive to smaller scale
lenses.
Because $K_L^{(i)}$ are smooth functions of $L$, we 
discretize $C_L^\PP$ into bins of width $\Delta L = 5$ 
(see \cite{Motloch:2016zsl} for more details).   Unless otherwise specified, we retain $N=4$ PCs in order to
fully characterize all sources of lensing information (see  \S\ref{sec:pp}).

In addition to these lens
parameters we take the standard $\Lambda$CDM parameters for the unlensed CMB
spectra:  $\Omega_b h^2$, the physical baryon density;
$\Omega_c h^2$, the physical cold dark matter density; $n_s$, the tilt of the
scalar power spectrum; $\ln A_s$, its log amplitude at $k=0.05$ Mpc$^{-1}$; $\tau$ the optical depth through reionization, and $\theta_*$, the angular scale of the sound horizon
at recombination.
We use flat  priors with uninformative ranges on these parameters.

In the MCMC analyses with $\Theta^{(i)}$, these cosmological parameters 
are still present and determine the ``unlensed" CMB fluctuations at recombination as well as the
background expansion.  However, they do not affect the lens potential as they would in the standard
analysis. We will refer to these parameters collectively as $\tilde \theta_A$, where  the
tilde is to remind the reader that the gravitational lens potential is not changed by
these parameters.

The fiducial cosmological model used to calculate $C_{L, \mathrm{fid}}^\PP$ in Eq.~\eqref{definition} is taken from 
the best fit flat $\Lambda$CDM cosmological model, determined from \verb|TT+lowTEB|
likelihoods assuming no primordial tensor modes and minimal
mass neutrinos ($\sum m_\nu = 60\, \mathrm{meV}$).
To reflect the latest results on the optical depth to recombination $\tau$
\cite{Adam:2016hgk}, we set $\tau$ to the value from that work and decrease $A_s$
to keep $A_s e^{-2\tau}$ constant.  A lower $A_s$ tends to exacerbate the preference for
anomalously high lensing in the temperature power spectrum
within the $\Lambda$CDM context but here serves only as the baseline fiducial model against
which to define $\Theta^{(i)}$.
 Values of the corresponding cosmological
parameters are listed in Table~\ref{tab:fiducial}.

\begin{table}
\caption{$\Lambda$CDM parameters and their fiducial values 
for the lens PC construction\footnote{In $\Lambda$CDM, these parameters also imply a Hubble constant
of $h=0.6733$.}.
}
\label{tab:fiducial}
\begin{tabular}{c|c}
\hline\hline
Parameter & Fiducial value\\
\hline
100 $\theta_*$ & 1.041 \\
$\Omega_c h^2$ & $0.1197$ \\
$\Omega_b h^2$ & $0.02223$ \\
$n_s$ & $0.9658$\\
$\ln(10^{10} A_s)$ & $3.049$  \\
$\tau$ &  $0.058$\\
\hline\hline
\end{tabular}
\end{table}

In models beyond $\Lambda$CDM, changes in the integrated Sachs-Wolfe (ISW) effect would
typically affect data on the
largest scales. In this work we are interested only in lensing-like effects and leave the ISW contribution at its $\Lambda$CDM value.

{Because the cosmic variance fluctuations of the lensing PCs $\Theta^{(i)}$ are by
at least factor eight smaller than the precision with which Planck power spectra can
measure $\Theta^{(i)}$, it is not necessary to consider the lensing-induced covariance terms
discussed in \cite{BenoitLevy:2012va, Motloch:2016zsl} in our analysis.  }

\section{Model-independent lensing constraints}
\label{sec:model_indept_lensing_constraints}

Measuring the lens principal components from the  Planck temperature, polarization and
lensing reconstruction power spectra 
provides a new means of extracting and comparing the various sources of lens
information in the CMB. This comparison presents a direct and model-independent consistency test of
the lensing information \cite{Motloch:2016zsl} and in the $\Lambda$CDM model context 
enables an  internal consistency {check} of the 
$\Lambda$CDM parameters inferred from CMB power {spectra} information from recombination
and from lens information. This is particularly
relevant given the known
preference for excess lensing in the Planck temperature data \cite{Ade:2013zuv,Ade:2015xua}.

We start by characterizing the information in the lensing reconstruction data, which  also determines the number of
lens principal components required for our comparative analysis.   We then discuss constraints on 
$\Theta^{(i)}$ from the temperature and polarization power spectra and focus on the two leading principal
components.   We also derive $\Theta^{(i)}$ from the recombination or unlensed information in the power spectrum
in the $\Lambda$CDM model as a check of its internal consistency.
We finish by commenting on robustness of our results with respect to various
analysis choices.

\subsection{Reconstruction constraints}
\label{sec:pp}

The principal component decomposition of the lens power spectrum described in \S\ref{sec:pc_intro} is optimized for the temperature
power spectrum analysis but can also be used to analyze the reconstruction data.
Even though we will be mainly interested in the first two components for comparisons with the other analyses, it is important to
retain a sufficient number of higher components so that their marginalization does not affect the lower ones.  

In practice we choose the number of principal components according to whether the data can 
constrain them better than a weak theoretical prior.  We choose flat tophat priors on $\Theta^{(i)}$
within a range where the variation in $C_L^\PP$ is within a factor of 1.5 of $C_{L,\rm fid}^\PP$.  These weak priors are meant to eliminate cases that would be in conflict
with other measurements of large scale structure or imply unphysically large amplitude high frequency
features in $C_L^\PP$.   

\begin{figure}
\center
\includegraphics[width = 0.49 \textwidth]{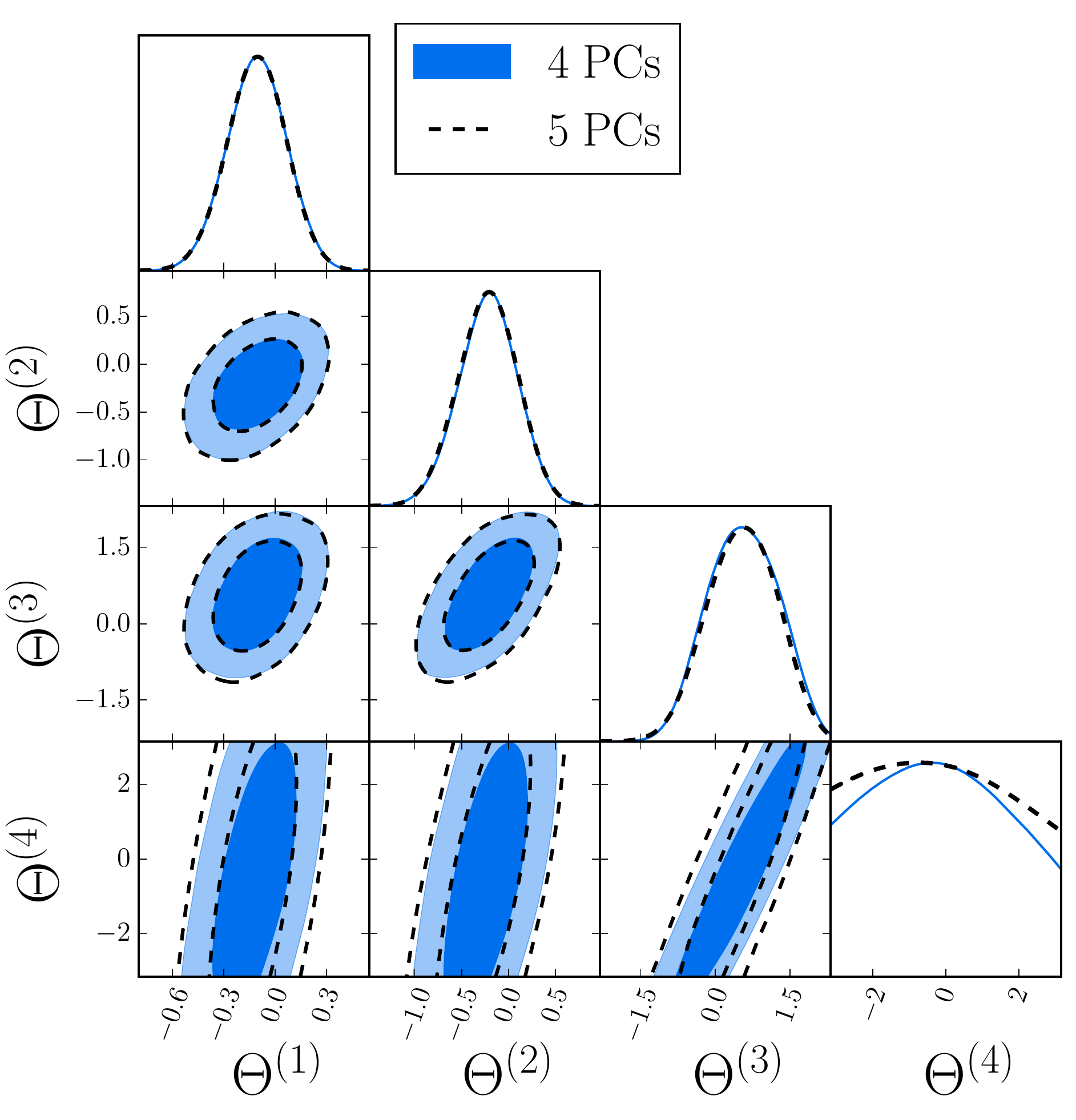}
\cprotect\caption{Lens reconstruction constraints from  \verb|PP| 
 on lens PCs (68\% and 95\% CL).  The  analysis with the fiducial four PCs (blue solid) and with an additional fifth PC marginalized (dashed) give nearly identical results for the first two PCs.
Higher PCs are fixed to zero, their fiducial value. }  
 \label{fig:lensonly}
\end{figure}

In Fig.~\ref{fig:lensonly} we show constraints from the lensing reconstruction
likelihood \verb|PP| on the first four $\Theta^{(i)}$ for the cases
where we allow four or five lens PCs to vary.   {For $\Theta^{(3)}$ and $\Theta^{(4)}$}
the edges of the box represent the prior and so the 4th component is nearly prior limited.
Correspondingly, the addition of $\Theta^{(5)}$ does not significantly affect constraints
on
$\Theta^{(1)}$ and $\Theta^{(2)}$ which will be important for evaluation of the tensions
in the data.    We therefore standardize on four lens PCs unless otherwise specified, with
higher lens PCs set to zero, which is their fiducial value.

\begin{figure}
\center
\includegraphics[width = 0.49 \textwidth]{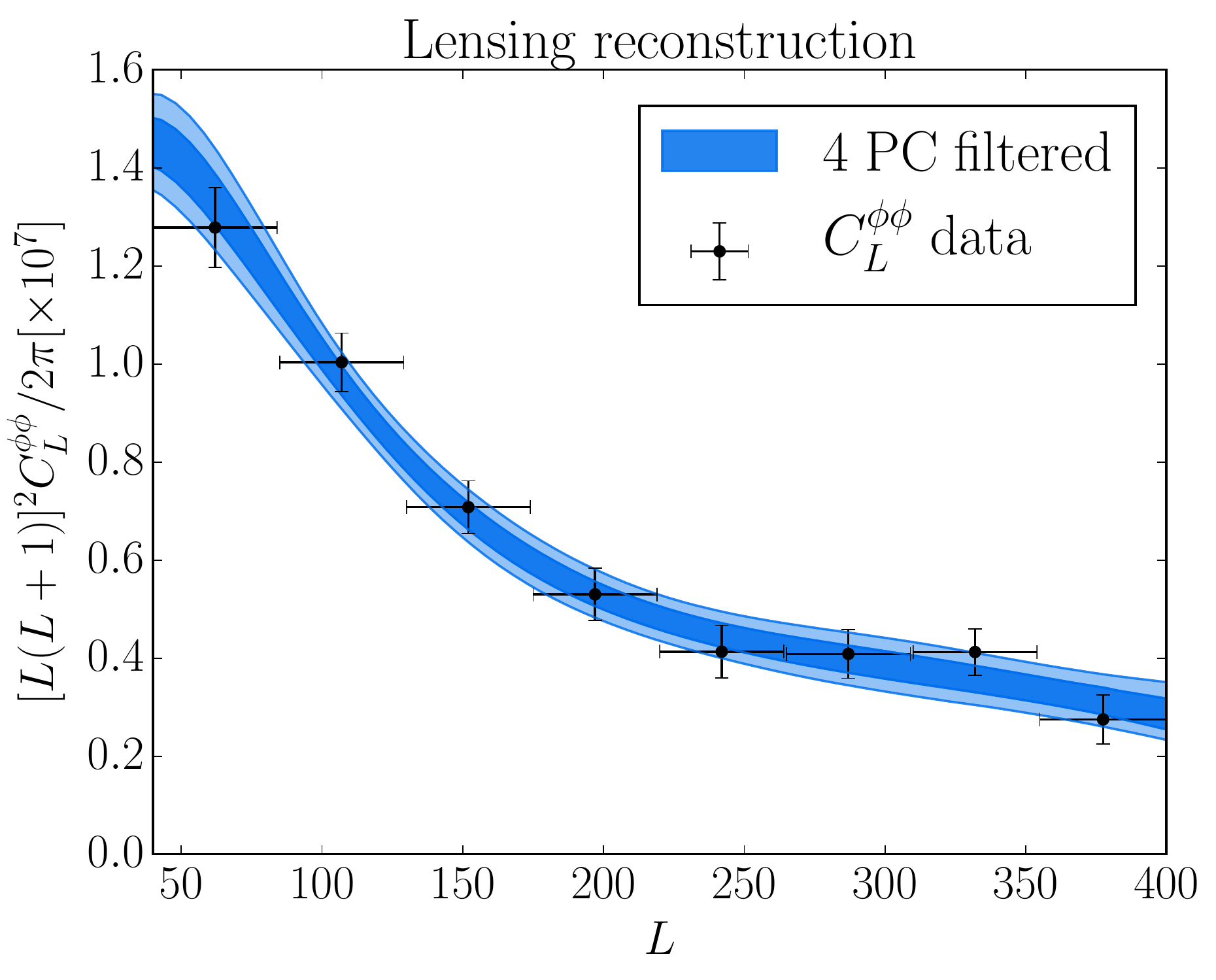}
\cprotect\caption{Lens reconstruction constraints from \verb|PP| on the lens power spectrum filtered through the 4 PC analysis $C_{L, \rm filt}^\PP$ (blue, 68\% and 95\% CL).
The points correspond to the measured Planck values included in the \verb|PP|
likelihood.  Although the points are only weakly correlated, PC filtering through
Eq.~(\ref{filteredpower}) utilizes all data points for 
each multipole leading to a smoother but correlated constraint.
}
\label{fig:cpp_posterior_just_pp}
\end{figure}

Another way to visualize why 4 PCs suffices is to  construct the lens power spectrum out of them as
\be
\label{filteredpower}
	C_{L,\rm filt} ^\PP  = C_{L, \mathrm{fid}}^\PP \exp\(\sum_{i = 1}^4 K^{(i)}_L\, \Theta^{(i)} \) .
\ee
and compare it to the lens reconstruction data itself.
In Fig.~\ref{fig:cpp_posterior_just_pp}, we show this comparison.  The 4 PC  construction represents smooth deviations
that are allowed by the data.  
Fluctuations that are not represented by the functional form
of the PCs shown in Fig.~\ref{fig:pcs} are not captured by the construction, for example the fluctuation in the data around $L=330$.
Thus the PC construction does not represent direct, but rather filtered, constraints on $C_L^\PP$.   To compare PC constraints from 
other sources to the lens reconstruction constraints, it is important to compare their
implications for  $C_{L,\rm filt}^\PP$ rather than $C_L^\PP$ directly.
This PC filter has the benefit of producing smooth functional constraints
utilizing the full data set at the expense of highly correlating constraints at 
different multipoles.

\begin{figure}
\center
\includegraphics[width = 0.49 \textwidth]{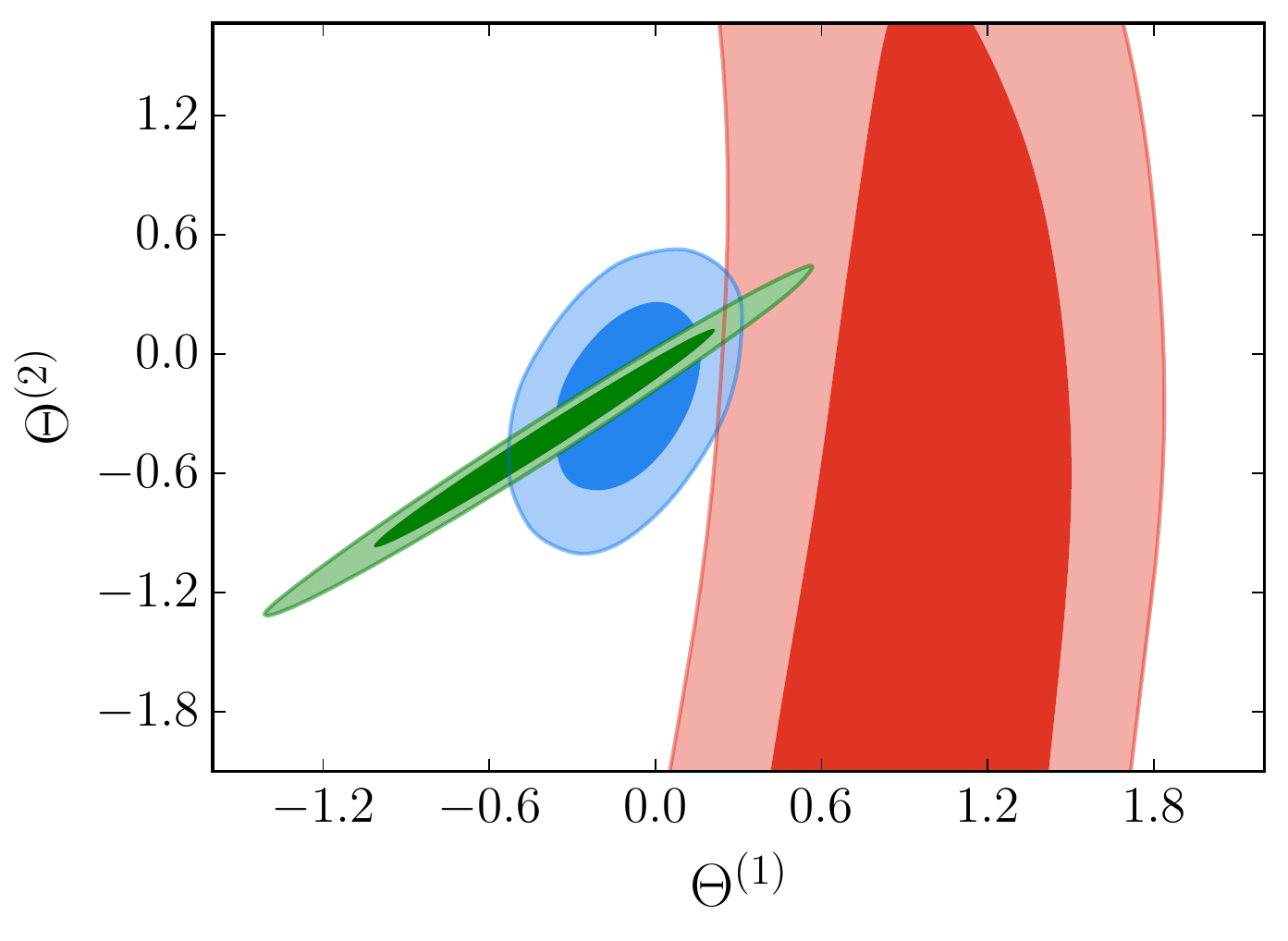}
\cprotect\caption{CMB power spectrum constraints on lens PCs $\Theta^{(1)}$ and
$\Theta^{(2)}$ 
from \verb|TT+lowTEB| (red,  68\% and 95\% CL) 
compared with lens reconstruction \verb|PP| from Fig.~\ref{fig:lensonly} (blue) 
and $\Lambda$CDM predictions based on unlensed parameters $\tilde \theta_A$ from \verb|TT+lowTEB| (green).  The fiducial 4 PC analysis is used in all cases.
}
\label{fig:tt_priored}
\end{figure}

\subsection{Temperature constraints}
\label{sec:tt}

We analyze the \verb|TT+lowTEB| likelihood for 4 PCs with the weak theoretical priors
discussed in the previous section.  
To focus on the region consistent with lens reconstruction, we impose an
additional data-driven prior. As shown in the previous section, lensing reconstruction constrains  $\Theta^{(1)}$ and $\Theta^{(2)}$
significantly better than the theoretical prior from the previous section, we thus
consider restricting these two variables further. As we will see, $\Theta^{(1)}$ drives
the tension between reconstruction and temperature constraints; for this reason we do not strengthen the prior on it. On the other
hand, we restrict $\Theta^{(2)}$ to lie within six standard deviations from the mean value
from the reconstruction analysis that considers 4 PCs. We retain this prior even for
analyses in which $\Theta^{(4)}$ is fixed to its fiducial value. We shall see that the
tension between power spectra and lensing reconstruction information on lensing 
is weaker
than 6$\sigma$ and so this prior does not artificially increase the tension. 
It therefore just excludes the parameter space that would
be grossly ruled out by reconstruction data and is used mostly for visualization
purposes.

\begin{figure}
\center
\includegraphics[width = 0.49 \textwidth]{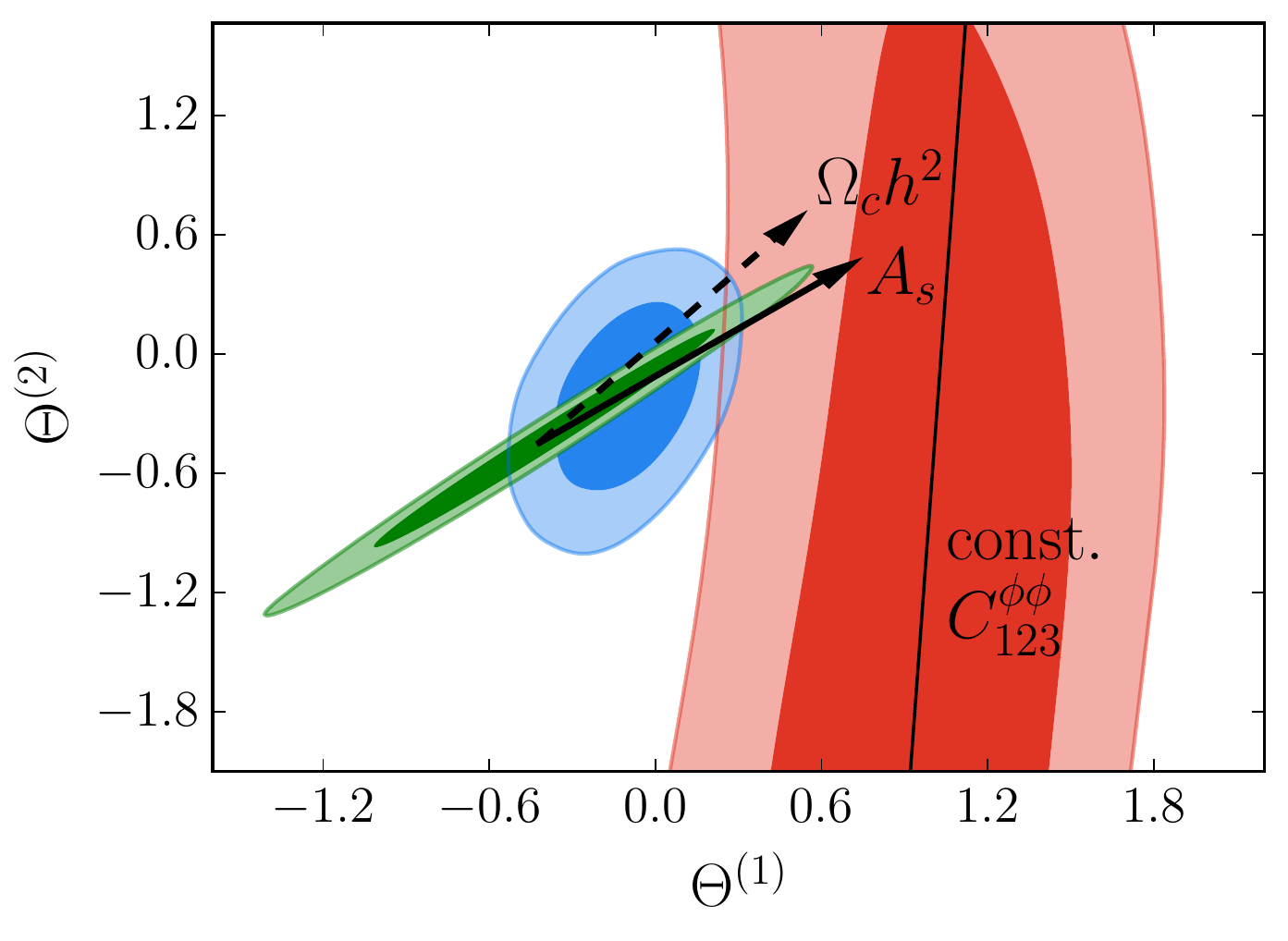}
\cprotect\caption{Physical interpretation of constrained directions from Fig.~\ref{fig:tt_priored}. 
The line approximates the degeneracy direction of the \verb|TT+lowTEB| (red) contour and
correspond to a line of constant $C_{123}^\PP$.
Arrows shows changes in
$C_L^\PP$ caused by increasing $A_s$ (solid) and $\Omega_c h^2$ (dashed) in $\Lambda$CDM
while keeping the other parameters in Tab.~\ref{tab:fiducial} fixed.
}
\label{fig:directions}
\end{figure}

As expected, \verb|TT+lowTEB| {data constrain} mainly one principal
component with  $\Theta^{(2)}$  limited by the priors and
$\Theta^{(3)}$ and $\Theta^{(4)}$ completely dominated by them.
In Fig.~\ref{fig:tt_priored} we show the constraints (red contours)  in the
$\Theta^{(1)}-\Theta^{(2)}$ plane out to the edge of the $\Theta^{(2)}$ prior.  Because
the PCs were constructed from a Fisher forecast, 
the constrained direction
is nearly but not perfectly aligned with $\Theta^{(1)}$, leaving a slight correlation between the
two parameters.   
In Fig.~\ref{fig:directions} we show that the degenerate direction for the  \verb|TT+lowTEB|  analysis 
corresponds approximately to 
constant 
$C_{123}^\PP$, whereas  contours of constant $\Theta^{(1)}$ correspond approximately to
 $C_{127}^\PP$ as determined by the zero crossing of $K_L^{(2)}$ in Fig.~\ref{fig:pcs}.   

For comparison we in Fig.~\ref{fig:tt_priored} also show the constraints from lens reconstruction (blue contours).
The two constraints are in tension with each other in that the two contours only overlap in their 95\% CL regions.
Moreover, this tension is model independent: 
no change in the shape of $C_L^\PP$ allowed by the 4 PCs can resolve it.

\begin{figure}
\center
\includegraphics[width = 0.49 \textwidth]{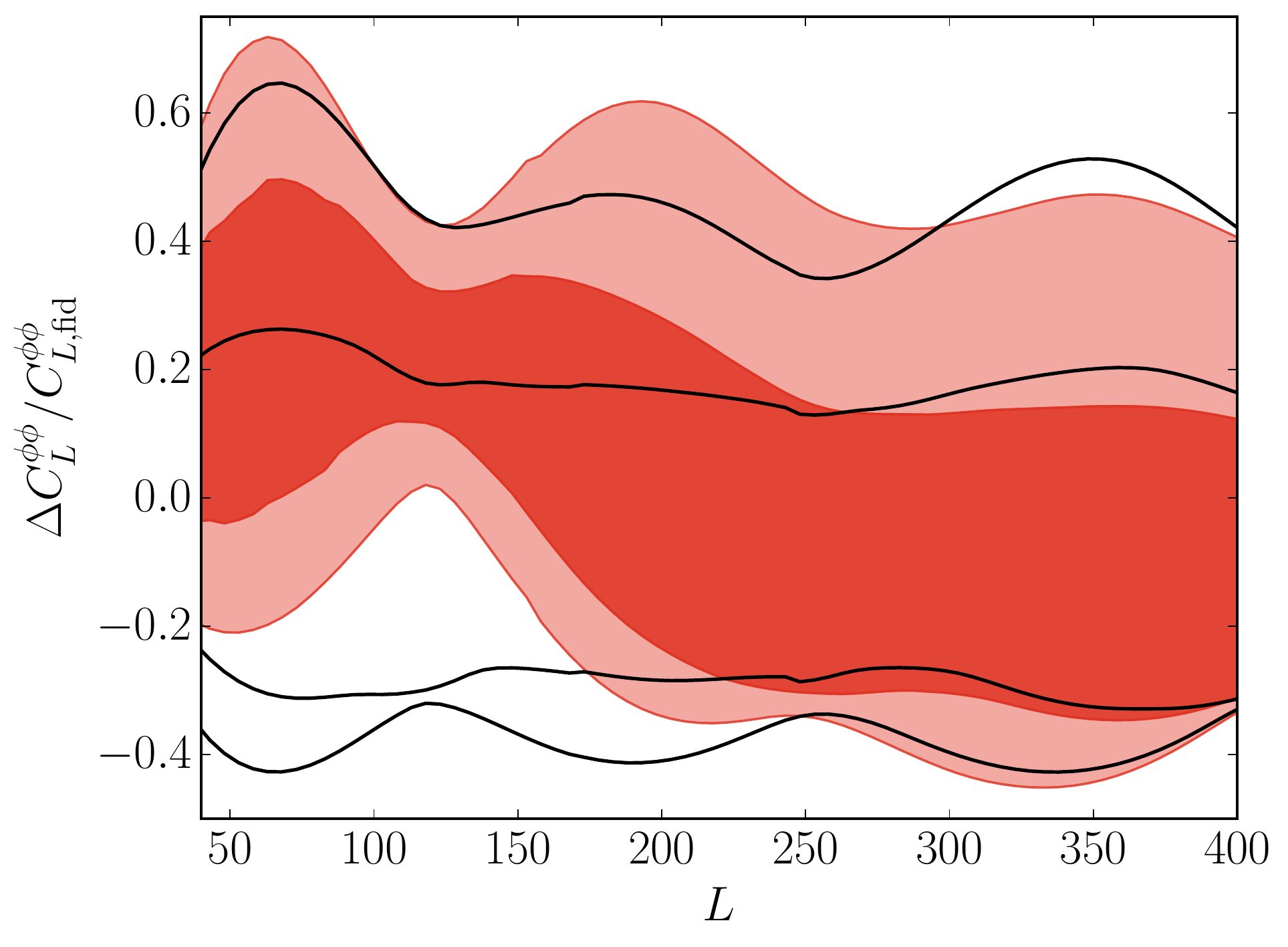}
\cprotect\caption{CMB power spectrum posterior constraints on  fractional deviations in the 4 PC
filtered lens power spectrum from the fiducial model
 $\Delta C_{L}^\PP/ C_{L, \rm fid}^\PP$ (red, 68\% and 95\% CL).
Compared with the prior constraints (black, same CL), the data are 
informative mostly around $L\sim 120$, favoring high lensing power, and above $L \sim 250$ the prior dominates.   }
\label{fig:only_prior_comparison}
\end{figure}

We can also visualize the \verb|TT+lowTEB|  constraint 
on $C_L^\PP$ as filtered through  the first 4 PCs via
Eq.~\eqref{filteredpower};  for
the ease of comparison we show fractional difference from the fiducial model.   
   In Fig.~\ref{fig:only_prior_comparison} we show
that posterior constraints from the  \verb|TT+lowTEB| data are {tighter} than
the prior mainly around  $L \sim 120$ while at high $L\gtrsim 250$ the
constraints are prior dominated. 
Note that the prior is skew positive
allowing a tail to high $C_L^\PP$ where the probability drops slowly.
The large values of $\Theta^{(1)}$ that the data prefer can therefore easily
push the $95\%$ CL region of the posterior beyond that of the prior,
especially around $L \sim 200$.

\begin{figure}
\center
\includegraphics[width = 0.49 \textwidth]{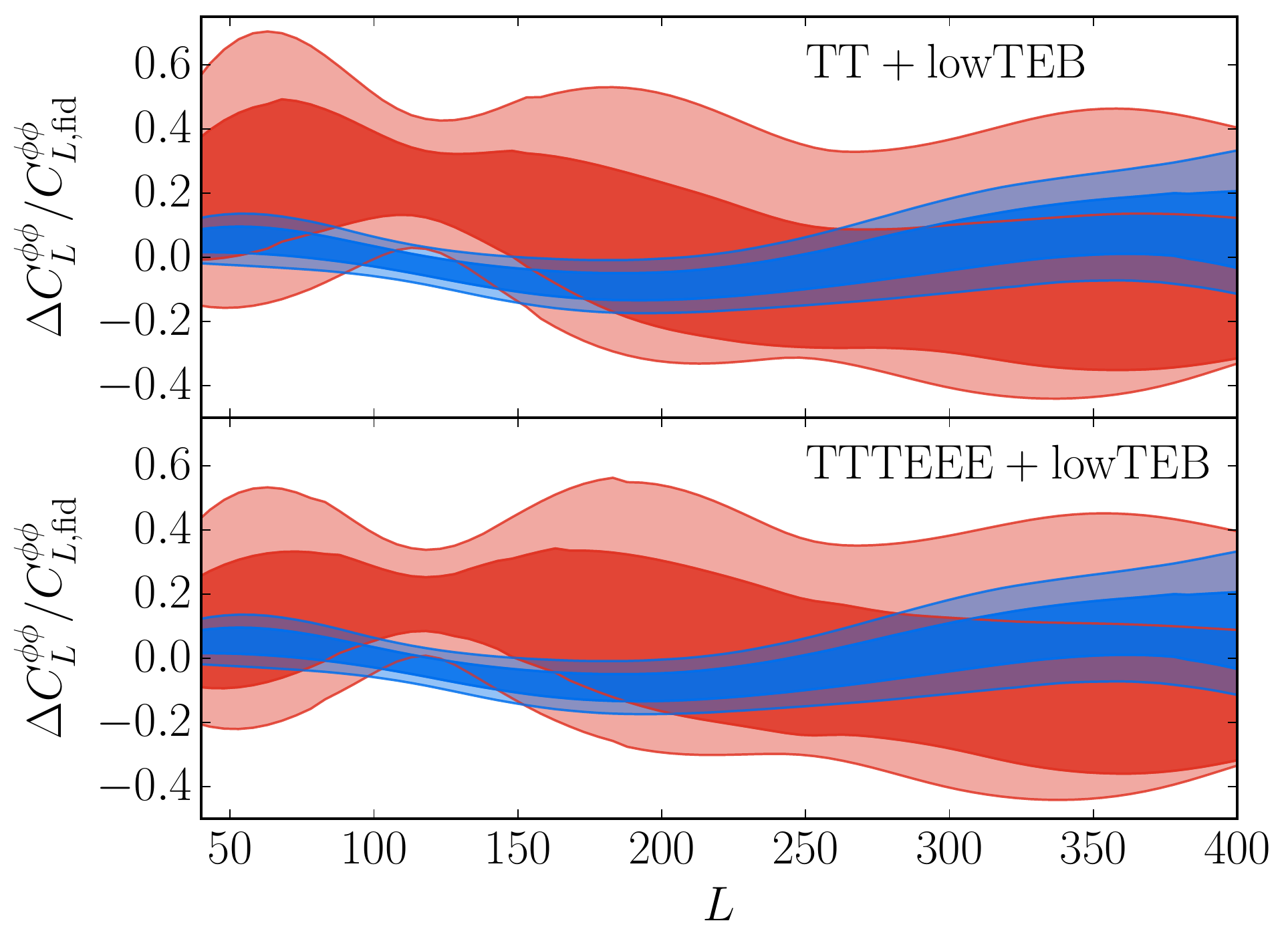}
\cprotect\caption{CMB power spectrum constraints on the filtered
 $\Delta C_{L}^\PP/ C_{L, \rm fid}^\PP$ as in Fig.~\ref{fig:only_prior_comparison}  compared with that of lens reconstruction
 from Fig.~\ref{fig:cpp_posterior_just_pp} (blue).  Top panel shows  constraints from
\verb|TT+lowTEB|  and bottom panel  from \verb|TTTEEE+lowTEB| which adds
high-$\ell$
polarization.    }
\label{fig:cpp_posterior}
\end{figure}

In Fig.~\ref{fig:cpp_posterior} (top panel) we compare these posterior constraints on $C_{L,\mathrm{filt}}^\PP$ from the \verb|TT+lowTEB| to those from  \verb|PP|.  
The reconstruction data favors less power around  $L \sim 120$.  Although changes in the shape of $C_L^\PP$ can bring agreement
between the two away from this regime, tension remains there independent of the model.  
We have explicitly checked that {relaxing} the
theoretical priors does not aggravate this tension, despite the fact that the upper bounds
from the posterior and the prior 
around $L\sim 120$ coincide in Fig.~\ref{fig:only_prior_comparison}.

Within the $\Lambda$CDM we can further study the origin of this tension.   
From the same \verb|TT+lowTEB| analysis, we can 
predict $C_L^\PP$
from information in the unlensed CMB power spectra at
each sampled parameter point $\tilde
\theta_A$ under the $\Lambda$CDM assumption.  We can then translate this prediction into $\Theta^{(i)}$ by
 inverting 
 Eq.~\eqref{definition}.
These $\Lambda$CDM predictions, shown as the green contours in
Fig.~\ref{fig:tt_priored}, can be directly compared
with the lensing PC measurements
themselves.   Some tension between the red and green
 contours is visible, as they overlap only at the $\sim 95\%$ confidence levels; this is the lensing PC version of the
 well-known $A_L$ lensing anomaly in the high-$\ell$ TT data.

{Unlike  $A_L$, which only indirectly specifies $C_L^\PP$ by changing its
 amplitude relative to the $\Lambda$CDM prediction point by point in its parameter space, PCs directly change the
 amplitude and shape of $C_L^\PP$.   This allows us to more directly quantify the
 origin of lensing tension.}   It is straightforward to trace
back the origin of the $\Lambda$CDM degenerate direction in the $\Theta^{(1)}
-\Theta^{(2)}$ plane: arrows in Figure~\ref{fig:directions}  show how these two parameters
change when we increase values of $A_s$ and $\Omega_c h^2$ (at fixed $\theta_*$ and other
parameters; constructed from the partial derivatives listed in
Tab.~\ref{tab:lcdm_effects_on_lensing}), which within $\Lambda$CDM are the two parameters
with dominant effects on $C_L^\PP$. {Given current constraints on $\tau$, the degenerate
direction is mainly aligned with that of $A_s$, with a smaller contribution from $\Omega_c
h^2$}. On the other hand, the lensing PC constraints from \verb|TT+lowTEB| mainly
reflect $\Theta^{(1)}$ and {are} driven by the $L\sim
125$ region of the lens power spectrum that is best measured by  the TT spectrum.  Though
they are lens model independent, the lens reconstruction constraints are {in good
agreement} with the $\Lambda$CDM constraints in green.
Furthermore, the near alignment of the directions of the $\Lambda$CDM constraints and
the reconstruction constraints (blue) also suggest that the tension with power spectrum constraints (red)
cannot be significantly relieved by going beyond $\Lambda$CDM.

\begin{table}
\caption{Dependence of $C_L^\PP$ on selected $\Lambda$CDM parameters} 
\label{tab:lcdm_effects_on_lensing}
\begin{tabular}{ccc}
\hline\hline
& $\Theta^{(1)}$ & $\Theta^{(2)}$\\	
\hline
$\partial \Theta^{(i)}/\partial(\Omega_c h^2)$ & 82.0 & 99.1\\
$\partial \Theta^{(i)}/\partial \ln A_s$ & 7.45 & 5.95\\
\hline\hline
\end{tabular}
\end{table}

\begin{figure}[b]
\center
\includegraphics[width = 0.49 \textwidth]{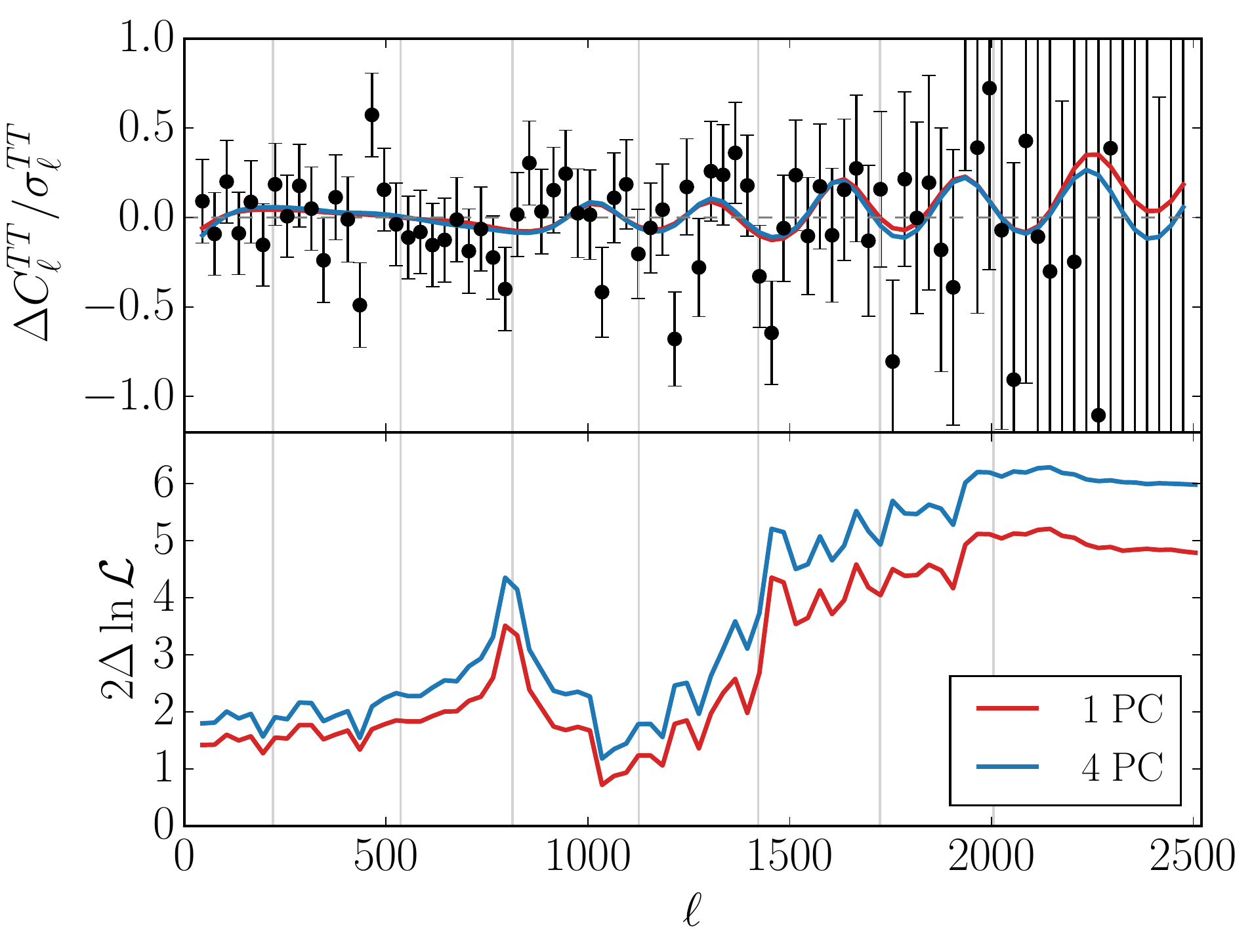}
\cprotect\caption{ Top: Residuals between Planck temperature power spectrum measurements
and the best fit $\Lambda$CDM model given the \verb|TT+lowTEB| likelihood (points,
scaled to cosmic variance errors per multipole $\sigma_\ell^{TT}$).  The blue (red) line shows the best fit
once we allow four (one) lensing PCs to vary, with fixed foregrounds. Gray vertical lines
show positions of the first seven acoustic peaks. Bottom:
improvement in the cumulative {$2\ln\mathcal{L}(\le \ell)$} over $\Lambda$CDM for
the same models showing that most of the improvement is from the first PC and corresponds
{to}
smoother acoustic peaks in the $\ell \sim 1250-1500$ range.
}
\label{fig:tt_residuals}
\end{figure}

\begin{figure*}
\center
\includegraphics[width = 0.99 \textwidth]{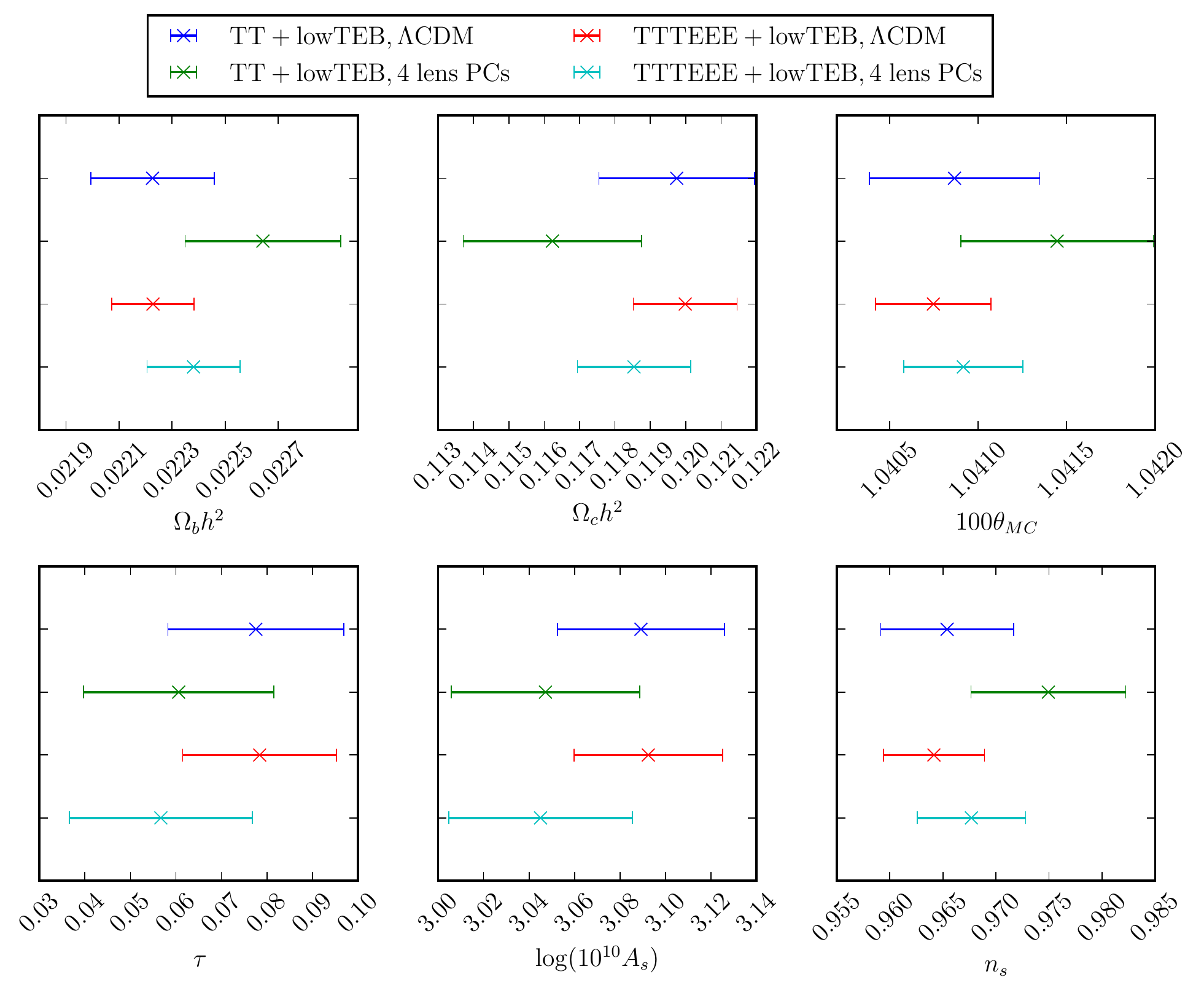}
\cprotect\caption{Cosmological parameters constraints from
\verb|TT+lowTEB| and \verb|TTTEEE+lowTEB| 
with the fiducial lensing 4 PC analysis (green, cyan) compared with $\Lambda$CDM
(blue, red).  The former correspond to  {constraints} on $\tilde \theta_A$ from the unlensed power
spectra. 
}
\label{fig:param_shifts}
\end{figure*}

We can also compare the temperature power spectrum of the maximum likelihood $\Lambda$CDM model with that of the maximum
likelihood model with lens PCs
to see what part of the temperature power spectrum data drives this
 preference for anomalous lensing; we show cases where either one or four lens PCs
are varying from their fiducial values.
 In Fig.~\ref{fig:tt_residuals}, we show the
 residuals relative to the best fit $\Lambda$CDM model scaled by the cosmic variance
 errors per multipole,
\begin{equation}
 \sigma_{\ell}^{TT} = \sqrt{\frac{2}{2\ell+1}} C_\ell^{TT},
\end{equation}
evaluated at the fiducial $\Lambda$CDM
model.  Notice the Planck data are binned and so the standard deviations of the data can be smaller
than $\sigma_\ell^{TT}$.
When searching for the best fit PC model, we fix the foreground
parameters  to their best fit $\Lambda$CDM values from the \verb|TT+lowTEB|
likelihood, for which the visualization of the Planck data points were
derived. 
In the lower panel of Fig.~\ref{fig:tt_residuals},  we then show the cumulative
improvement over $\Lambda$CDM in {$2\Delta \ln \mathcal{L}$}
of the
fit as a function of the maximum $\ell$.\footnote{Due to different binning schemes used by
the Planck collaboration for their best fit TT power spectrum and binned TT likelihood,
we use the \emph{unbinned} Planck TT likelihood to obtain this plot.}
The total reaches {$2\Delta\ln\mathcal{L}=6.0$} at the highest multipole employed in the analysis
when allowing four lens PCs to vary. 
As is visible from the figure, {$2\Delta \ln\mathcal{L} \approx 5$} of 6  comes from  the first
lens PC, in agreement with our previous statement
that majority of the lensing information in the temperature power spectra is well captured
by a single lensing component.

 The data points show oscillatory residuals with respect to the $\Lambda$CDM model  in the $\ell$ range 1250--1500 that 
indicate smoother acoustic oscillations (see also
\cite{Aghanim:2016sns,Obied:2017tpd}).   Correspondingly, the largest part of the
improvement  arises through fitting these residuals by increasing the smoothing due to lensing, though notable contributions come from the \verb|lowTEB| part of the likelihood.   The latter
is associated with the ability to lower TT power at $\ell \lesssim 30$.

These improvements allowed by releasing $C_L^\PP$ from its $\Lambda$CDM value also lead to
shifts in cosmological parameters; these shifts are summarized in
Figure~\ref{fig:param_shifts}, for the case where the foreground parameters are
again allowed to vary.
In $\Lambda$CDM, preference for fitting the oscillatory residuals 
pushes values of $\Omega_c h^2$ and $A_s$ up, which then forces other $\Lambda$CDM
parameters to compensate. In the PC case, lensing parameters
$\Theta^{(i)}$ play this role and allow $\Omega_c h^2, A_s$ to drop. 
This drop and the associated changes in other parameters allows for a lower low-$\ell$ TT power
with respect to the acoustic peaks and therefore  also allows a better fit to the anomalously 
low TT power at $\ell \lesssim 30$.
In models with the $\Lambda$CDM expansion history such a drop also
simultaneously raises $H_0$ to $\(69.1 \pm 1.2\)\,\mathrm{km/s/Mpc}$ and can help
relieve tension with the local distance ladder measurements \cite{Riess:2018aaa}.

Taken at face value, these mild tensions and their alleviation with lensing PC parameters
  would motivate explorations of additional physics
at low $z$ which modify the lens potential.  However, independent of the model for the lens potential, tension with
lensing reconstruction remains.

\subsection{Polarization constraints}
\label{sec:pol}

\begin{figure}
\center
\includegraphics[width = 0.49 \textwidth]{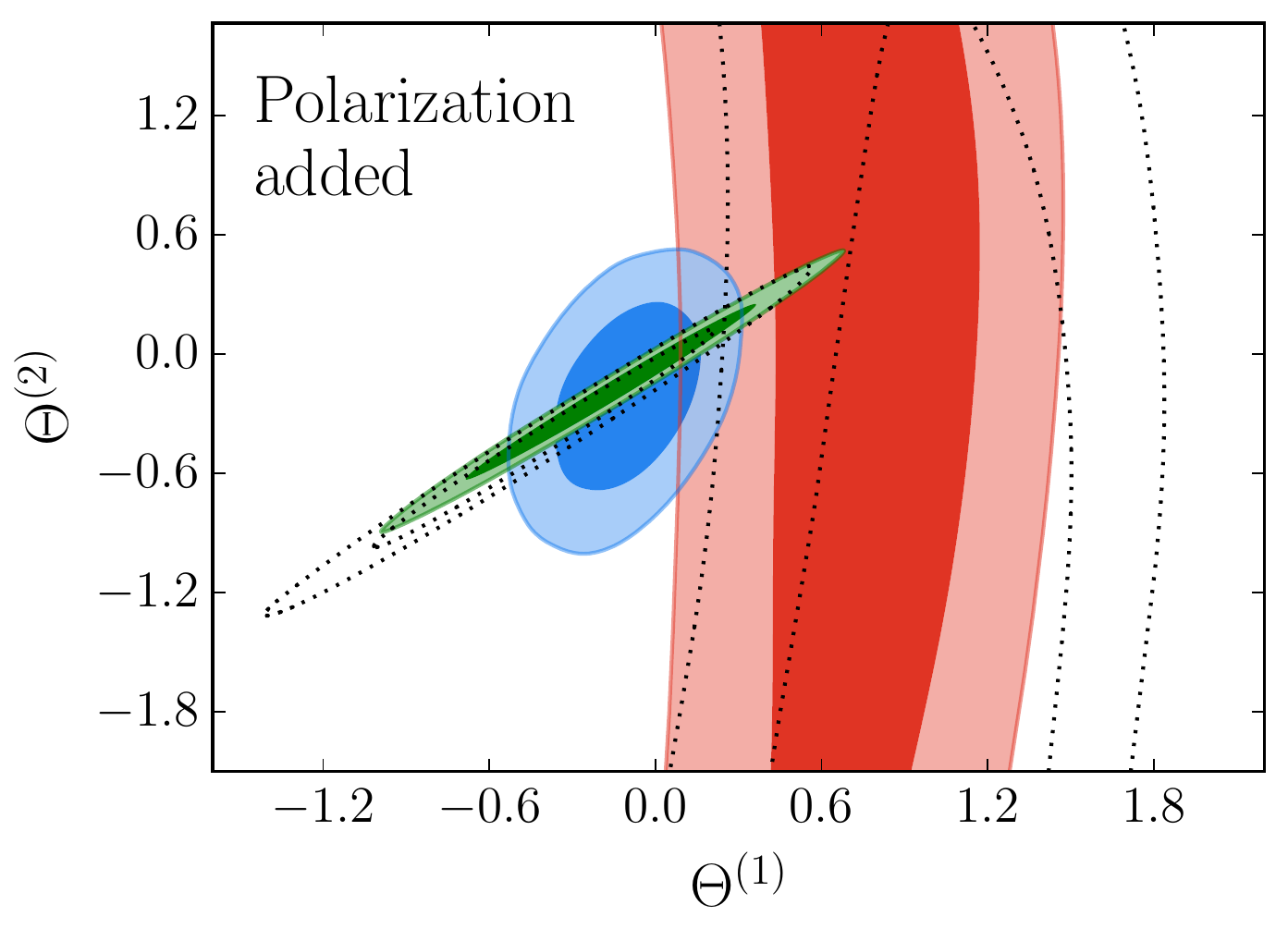}
\cprotect\caption{Impact of high-$\ell$ polarization on PC constraints
from Fig.~\ref{fig:tt_priored} (repeated with dotted contours for comparison).  While
tension between the \verb|TTTEEE+lowTEB| (red) and the $\Lambda$CDM results (green)
weakens slightly, its tension with \verb|PP| remains nearly the same due to its shift and reduced errors in the $\Theta^{(1)}$ direction.  }
\label{fig:pol_priored}
\end{figure}

Next, we add the high-$\ell$ polarization constraints using the
\verb|TTTEEE+lowTEB| likelihood; the various constraints on $\Theta^{(1,2)}$ are shown
in Fig.~\ref{fig:pol_priored}. The 2015 Planck polarization data is known to be subject to
systematics that make lensing conclusions unstable \cite{Ade:2015xua} and thus we consider
their addition separately.

The main change is a shift in the contours to lower values of $\Theta^{(1)}$ but with tighter
errors.   This shift is driven by the $C_\ell^{TE}$
data; lensing constraints from $C_\ell^{EE}$ are notably weaker and additionally favor
even more lensing than $C_\ell^{TT}$ does \cite{Ade:2015xua}.
With polarization, the tension between CMB power spectra and lens reconstruction constraints only mildly relaxes.
This is because of the combination of the shift and the smaller errors in Fig.~\ref{fig:pol_priored}.  In
Fig.~\ref{fig:cpp_posterior} (bottom), we also show the impact 
of adding polarization data on the filtered $C_L^\PP$ constraints.   Correspondingly, polarization data only
mildly decreases the significance of tension around $L \sim 125$.
 
The internal tension between  temperature-polarization power spectra and the $\Lambda$CDM
prediction in green relaxes somewhat more.  This is because polarization favors the
high $\Omega_c h^2$ values of the best fit $\Lambda$CDM model to
\verb|TT+lowTEB|
(as shown in Fig.~\ref{fig:param_shifts}) 
due to unusually strong TE constraints in the region around $\ell \sim 200$ \cite{Obied:2017tpd}.   This preference is in a region
that is relatively unaffected by lensing and so remains after releasing $C_L^\PP$.  
Raising $\Omega_c h^2$ in $\Lambda$CDM has the effect of increasing lensing making
the   temperature-polarization power spectra and $\Lambda$CDM somewhat more consistent.

\subsection{Robustness tests}
\label{sec:robustness}

\begin{figure}
\center
\includegraphics[width = 0.49 \textwidth]{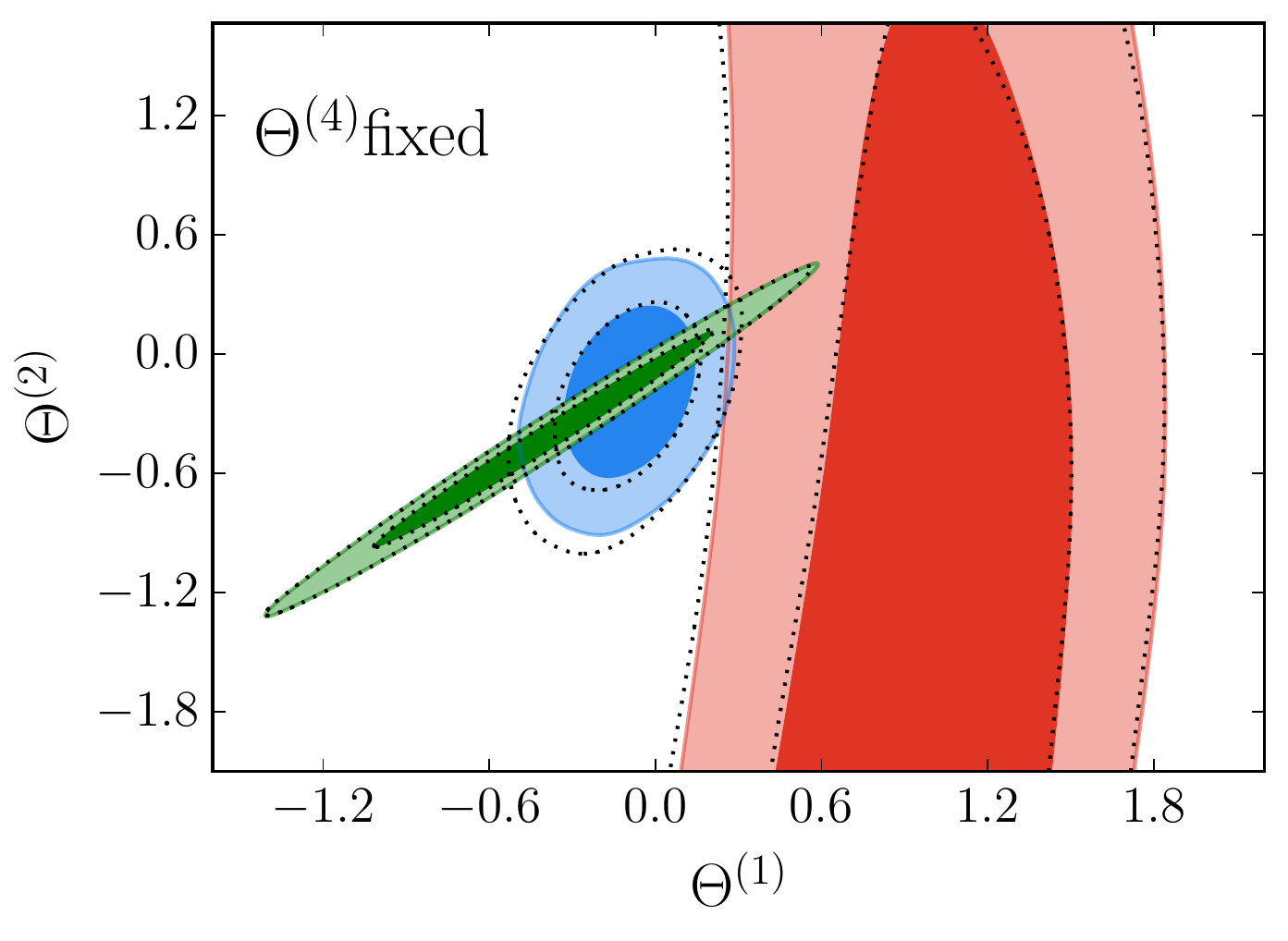}
\includegraphics[width = 0.49 \textwidth]{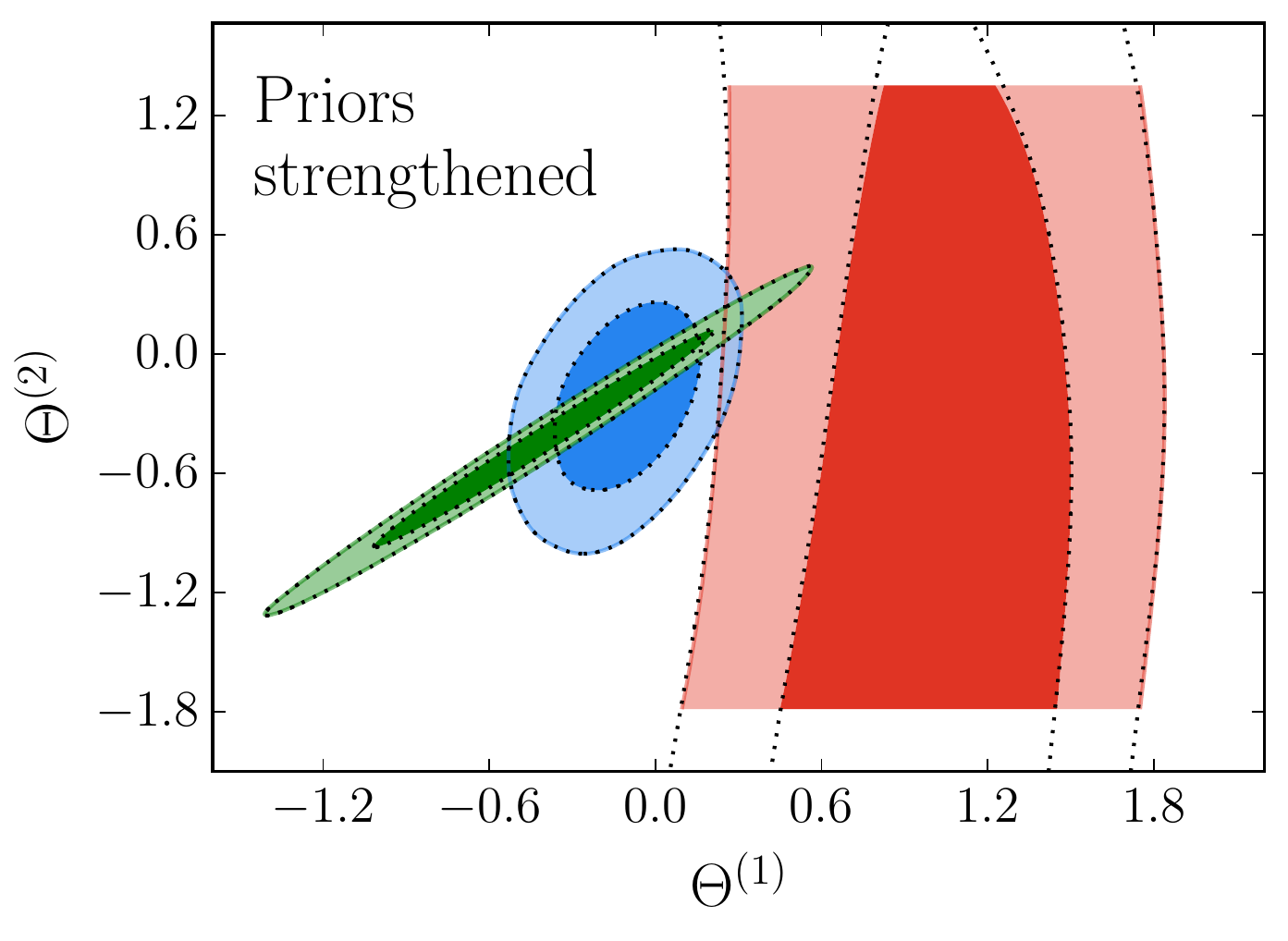}
\cprotect\caption{Robustness checks on PC constraints from  Fig.~\ref{fig:tt_priored}
(repeated with dotted contours for comparison).  Top: $\Theta^{(4)}$ fixed to its fiducial
value instead of marginalized over.  Bottom: {tighter} theoretical prior on $\Theta^{(i)}$
(see the text).  Neither change significantly impacts tension between the lensing
measurements.}
\label{fig:tt_priored_Npc}
\end{figure}

To check that the results are stable with respect to the considered number of lens
PCs, we repeat our analysis with $\Theta^{(4)}$ fixed to its fiducial value. In
Fig.~\ref{fig:tt_priored_Npc} (top) we show that this does not significantly alter the various
constraints on $\Theta^{(1)}, \Theta^{(2)}$ based on  \verb|TT+lowTEB| and
\verb|PP|. The same conclusion holds when polarization data are added.

We also repeat our analysis with a {tighter} theoretical prior (with four lens PCs
varying) -- for $\Theta^{(3)}$ and $\Theta^{(4)}$ we restrict the variation in
$C_L^\PP$ to be within a factor of 1.4 of $C_{L,\mathrm{fid}}^\PP$, instead of our default
1.5, while we demand $\Theta^{(2)}$ to be within five standard deviations from the
mean value determined from the  \verb|PP| likelihood.  In
Fig.~\ref{fig:tt_priored_Npc} (bottom) we show that impact of the theoretical prior
on the tension is negligible.

In our analysis we have so far fixed the unlensed CMB to 
the power spectra allowed by $\Lambda$CDM.
Given the lensing model-independence of the tension, it is also interesting to ask whether
additional physics at recombination can relax it.  
We can never completely eliminate this possibility for resolution of tension
with our methodology, as effects of this new physics might mimic lensing in the CMB
power spectrum while not affecting the higher point moments important in lens reconstruction.
On the other hand, we can show that
the additional physics cannot be simply a change in the  effective number of light
relativistic species $N_\mathrm{eff}$. When adding this parameter to the
unlensed parameters $\tilde \theta_A$ and marginalizing over it,
we find that the tension between high-$\ell$ TT and lensing reconstruction constraints and
the internal lensing tension are both still present and similarly significant.

\section{Significance of the tensions}
\label{sec:significance}

Having illustrated the existence of lensing tensions in the Planck CMB data, we now turn to quantifying
their significance. We start by defining a robust single statistic to compare between 
the various sources of lensing information.  We then discuss the significance of the   model-independent 
 tension between lensing
constraints from Planck temperature/polarization power spectra and lens
reconstruction. After that we focus on a special case of $\Lambda$CDM with a freely
floating amplitude of the lensing potential, which allows us to compare with previous
literature and address the significance of the internal tensions between $\Lambda$CDM
lensing constraints from within the CMB power spectra alone.

\subsection{Tension statistics}
\label{sec:tension_statistics}

In order to quantify the tension simply and cleanly, we seek to find a single auxiliary
parameter whose distribution reflects the best constraints and is as close
as possible to  Gaussian in each of the lensing measurements.   
For two such measurements, a natural tension statistic to use is the shift in the means $|\mu_{1}-\mu_{2}|$.  
To the extent that the parameter posteriors are Gaussian distributed, the shift
itself is predicted to be Gaussian distributed with
a variance that is the sum of the two variances, $\sigma^2=\sigma_1^2+\sigma_2^2$.
Therefore, the significance of
the measured shift in units of $\sigma$ is given by
\be	
\label{tension}
	T = \frac{\left|\mu_1 - \mu_2\right|}{\sqrt{\sigma_1^2 + \sigma_2^2}}.
\ee

\begin{figure}
\center
\includegraphics[width = 0.49 \textwidth]{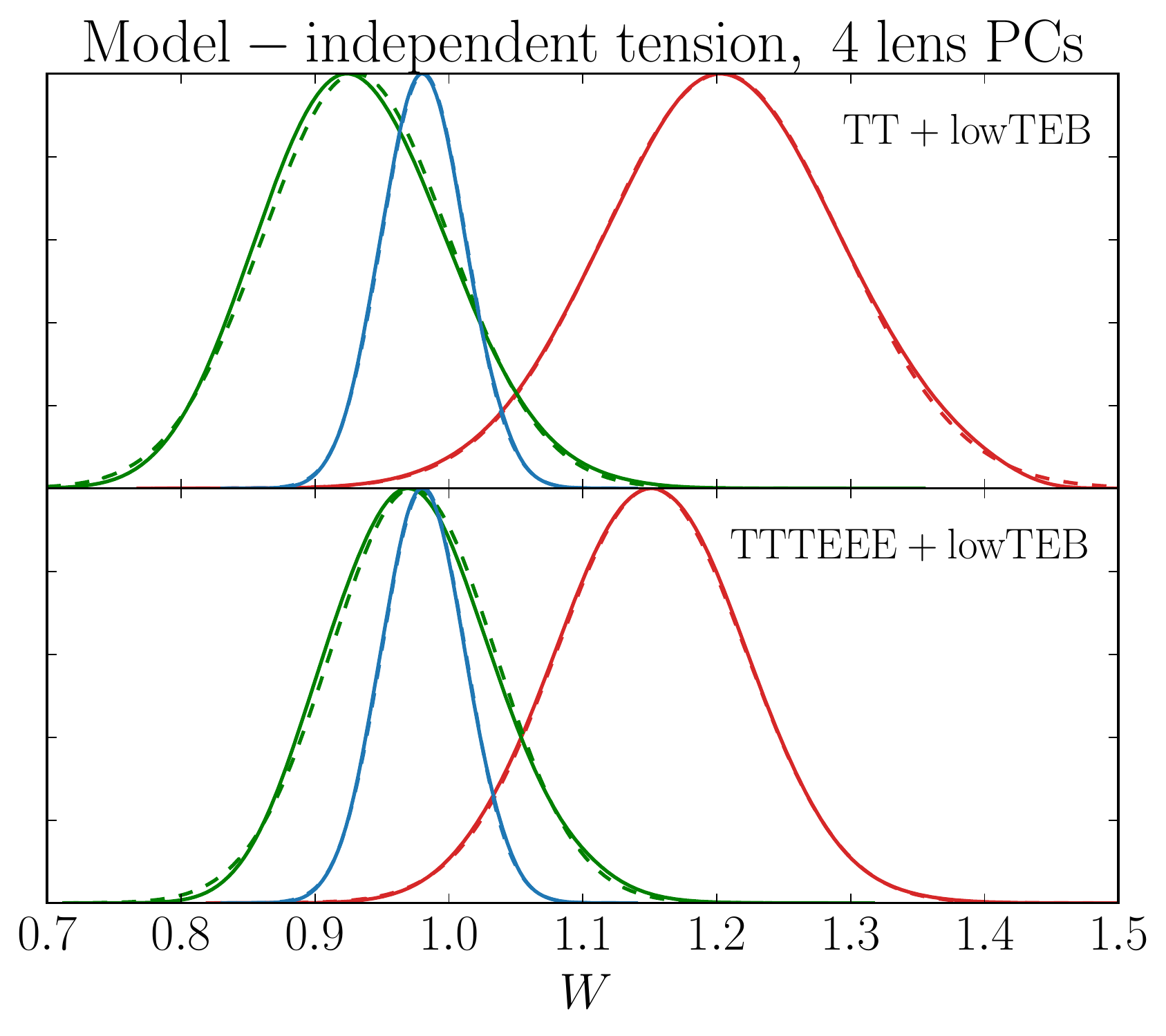}
\cprotect\caption{Posterior probability distribution for the lensing tension parameter $W$ as determined from the fiducial  4 PC analysis of lens reconstruction \verb|PP| (blue), 
and CMB power spectra
\verb|TT+lowTEB| (red, top) and \verb|TTTEEE+lowTEB| (red, bottom). In green we show
constraints on $W$ derived from $\tilde \theta_A$, obtained under the assumption of
$\Lambda$CDM from either \verb|TT+lowTEB| or \verb|TTTEEE+lowTEB|. Dashed lines show Gaussian
distributions with the same means and variances.}
\label{fig:tension_gaussianity}
\end{figure}

\begin{table}
\caption{Tension significances when comparing $W$ constraints from a reference data set to CMB power spectra constraints}
\label{tab:tensions}
\begin{tabular}{cccc}
\hline\hline ref.~data
& $C_L^\PP$ freedom & \verb|TT+lowTEB| &
\verb|TTTEEE+lowTEB| \\	
\hline
 \verb|PP| & 4 PCs & 2.4 & 2.2\\
 \verb|PP| & amplitude $\mathcal{A}$ & 2.4 & 2.4 \\
$\tilde \theta$ unlensed & amplitude $\mathcal{A}$ & 2.4 & 2.1\\
\hline\hline
\end{tabular}
\end{table}

To choose the parameter itself, note that the main source of tension is 
the first principal component  $\Theta^{(1)}$ (see  Fig.~\ref{fig:tt_priored}).   
However, 
to have the posterior distributions well approximated by Gaussian distributions, we 
instead choose
\be
\label{W}
	W \equiv \exp\(
	K_{123}^{(1)} \Theta^{(1)}
	\) .
\ee
$W$ is independent of the higher lens PCs, as these are not constrained by
the temperature and polarization power spectra and thus do not add to the tension.

The scaling factor $K_{123}^{(1)}$ makes $W$ the ratio of the
 1 PC filtered and fiducial $C_L^\PP$ evaluated at $L=123$  (see Eq.~(\ref{filteredpower})).
 As we will see, the main benefit of $W$, or in general a smoothly filtered version of $C_L^\PP$,
is that it represents a weighted average in $L$ of the data even
though it appears to be evaluated at a fixed $L$. As such it employs the constraining
power of the full range of the data. This leads to a powerful and robust tension statistic.

This should be contrasted with $C_{123}^\PP$ itself or more
generally the power spectrum at any single multipole $L$.   Its value  depends sensitively on the
higher PCs, which increasingly fit noise fluctuations, and so
represent an ineffective tension statistic when they are included. With our standard 4 PC
analysis, 
this is not a
significant problem for $C_{123}^\PP$ itself as we shall see in the next section, but by defining tension in $W$ we
make it robust to higher PCs as well and immune to reoptimizing the effective multipole for each case.

\subsection{Model-independent tension}

In Fig.~\ref{fig:tension_gaussianity}, we compare posterior
distributions for $W$ determined from  CMB power spectra through the \verb|TT+lowTEB|, \verb|TTTEEE+lowTEB| 
likelihoods to that determined from reconstruction through the 
\verb|PP| likelihood; the two types of distributions overlap only in the tails.
Gaussians with the
same means and variances describe even these overlap regions accurately, which justifies
the use of the tension statistic $T$.  
The tension between \verb|TT+lowTEB| and \verb|PP| determinations of $W$ is
significant at 2.4$\sigma$; adding polarization data decreases the tension to 2.2$\sigma$.
We summarize significance of various tensions in Table~\ref{tab:tensions}.

\begin{figure}
\center
\includegraphics[width = 0.49 \textwidth]{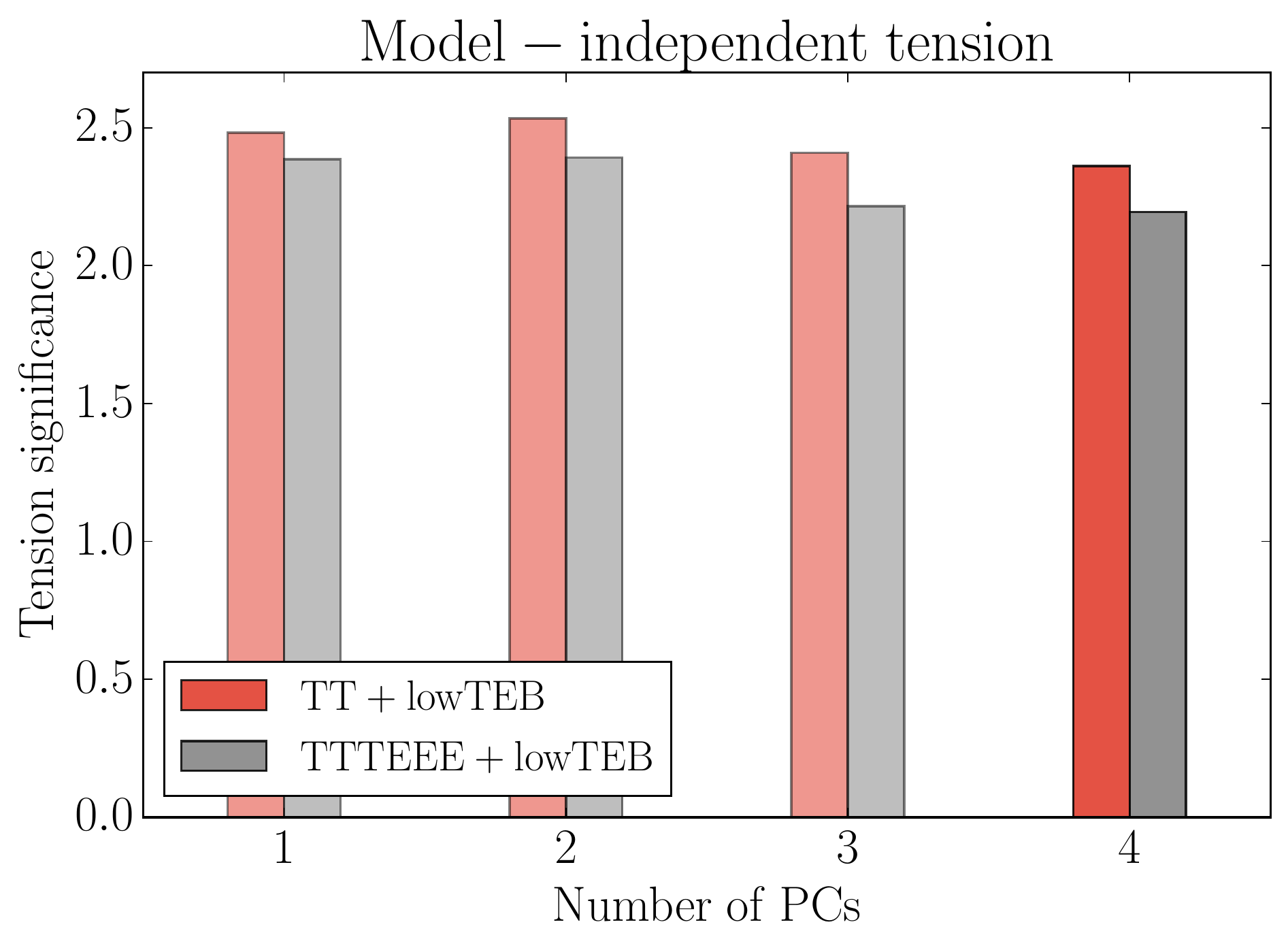}
\cprotect\caption{Significance of the {model-independent} tension between \verb|PP|
and \verb|TT+lowTEB| (red) or \verb|TTTEEE+lowTEB| (gray) determinations of $W$, as a
function of the number of the lensing PCs which are allowed to vary.   The tension
significance is measured in units of $\sigma$, the expected {root mean square} of
the distance between the means. {Our default result that uses the 4 PCs is
highlighted.}
}
\label{fig:tension_on_npc}
\end{figure}

We now consider several robustness checks on this tension.
Using the
foreground-marginalized high-$\ell$ TT likelihood \verb|liteTT| instead of
\verb|TT| {in
the analysis leads to the same tension significance of 2.4$\sigma$.}
When the data-driven prior on $\Theta^{(2)}$ of six standard deviations from the \verb|PP| 
constraint is dropped, the tensions relax both by  about 0.1$\sigma$.
This is caused by the small curvature of the posterior in the $\Theta^{(1)} -
\Theta^{(2)}$ plane, visible for example in the red contour in
Fig.~\ref{fig:pol_priored}, which leads to an increased overlap with the lensing
reconstruction constraints after the projection onto $W$.

\begin{figure}
\center
\includegraphics[width = 0.49 \textwidth]{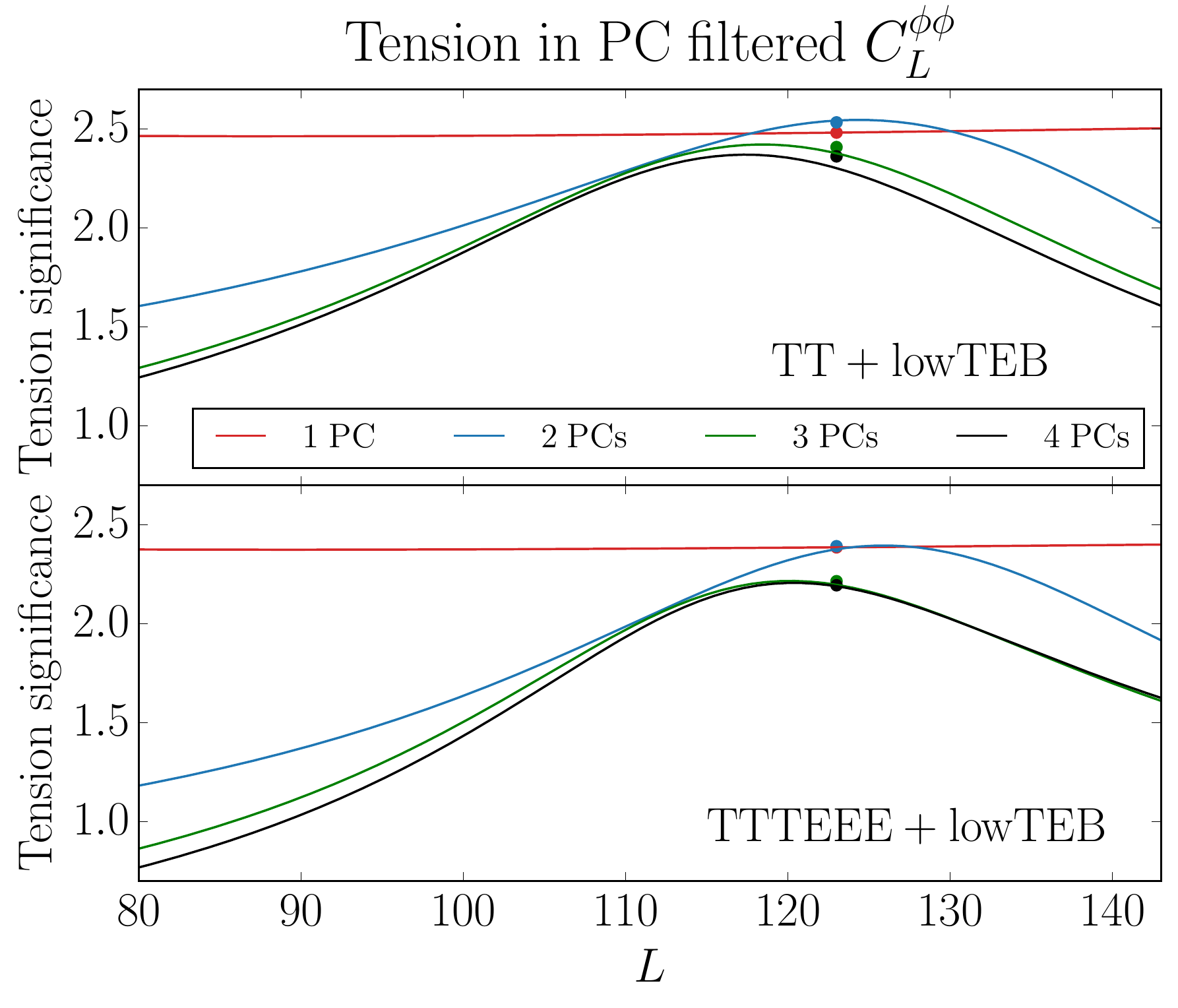}
\cprotect\caption{Tension significance for the PC filtered $C_L^\PP$ from \verb|PP| and \verb|TT+lowTEB| (top) or
\verb|TTTEEE+lowTEB|
(bottom) for various values of $L$ and number of lensing PCs which are allowed to vary. The
points represent significance of the {model-independent} tensions based on $W$;
notice that for one lens PC the two tension statistics are identical at $L=123$ by construction but
that $C_L^\PP$ at other values can substantially underestimate tension. 
}
\label{fig:tension_cpp}
\end{figure}

Next we consider robustness to the constraint on the reionization optical depth $\tau$.
The upcoming final release of Planck data is expected to improve and potentially change these constraints. 
Furthermore, constraints on $\tau$ depend on the form assumed for the ionization history
that is taken
to be step-like in the standard analysis \cite{Hu:2003gh,Heinrich:2016ojb}.
By isolating the information on the lens power spectrum itself, our tension statistic should be immune to such changes.
To quantify the
impact of possible future changes in the likelihood, we reevaluate
the tension statistic 
where instead of using \verb|TT+lowTEB| we 
constrain $W$ using \verb|TT+lowT|, together with a
$\tau$
prior of width 0.02, centered on either 0.04, 0.06 or 0.08. 
In all three cases the tension changes by less than 0.02$\sigma$ from the
original result obtained using \verb|TT+lowTEB|. Our conclusions are thus robust to 
 the low-$\ell$ polarization data and
likelihoods. 
Contrast this with the scaled $\Lambda$CDM approach where $\tau$ changes
the lens power spectrum  against which $A_L$ is measured from  the
temperature power spectrum and lens reconstruction data respectively, leading to
sensitivity of $A_L$ constraints to reionization assumptions.

The significance of the tension  between
\verb|TT+lowTEB| (or \verb|TTTEEE+lowTEB|) and \verb|PP| determinations of $W$ does not
notably change when we decrease the freedom in varying $C_L^\PP$ by retaining a smaller
number of lens PCs in the analysis (see Fig.~\ref{fig:tension_on_npc}).

Finally, it is possible to demonstrate why $W$ is more robust than $C_L^\PP$ at some
$L$ that has not been specifically optimized for the model-independent lensing test.
First, we can take the full 4 PC filtered construction of $C_L^\PP$ depicted in
Fig.~\ref{fig:cpp_posterior}. 
We show the resulting tension as a function of $L$ in
Fig.~\ref{fig:tension_cpp} ({black curves}) between
\verb|PP| and \verb|TT+lowTEB| (or \verb|TTTEEE+lowTEB|). 
The tension $T$ in $C_{L}^\PP$ constraints at $L \sim 120$ is similar to that in
$W$.   On the other hand, choosing other values of $L$ could
substantially degrade the ability to identify tension in these cases where the shape of $C_L^\PP$ is
allowed to vary.

Likewise, an alternate choice of $L$ can make the tension statistic more dependent on the number of lens PCs
allowed to vary. In Fig.~\ref{fig:tension_cpp}, we also show the {results} of analyses
with 1,2 or 3 lens PCs allowed to vary,  where $C_L^\PP$ is filtered with the same number of PCs.
Again, away from $L\sim 120$ the tension significance can vary widely.

\subsection{{$\Lambda$CDM and amplitude changes }}

Besides the weak theoretical prior, the tension quoted in the previous section was
derived without any constraints on the shape of the gravitational lensing potential. By
allowing the largest possible freedom, it represents a
lower limit on the tension present in the data; particular models can restrict this
freedom and consequently lead to a
larger significance of the tension. As a simple example and to connect
with the earlier literature, we investigate here $\Lambda$CDM model with a freely floating
amplitude of $C_L^\PP$.

\begin{figure}
\center
\includegraphics[width = 0.49 \textwidth]{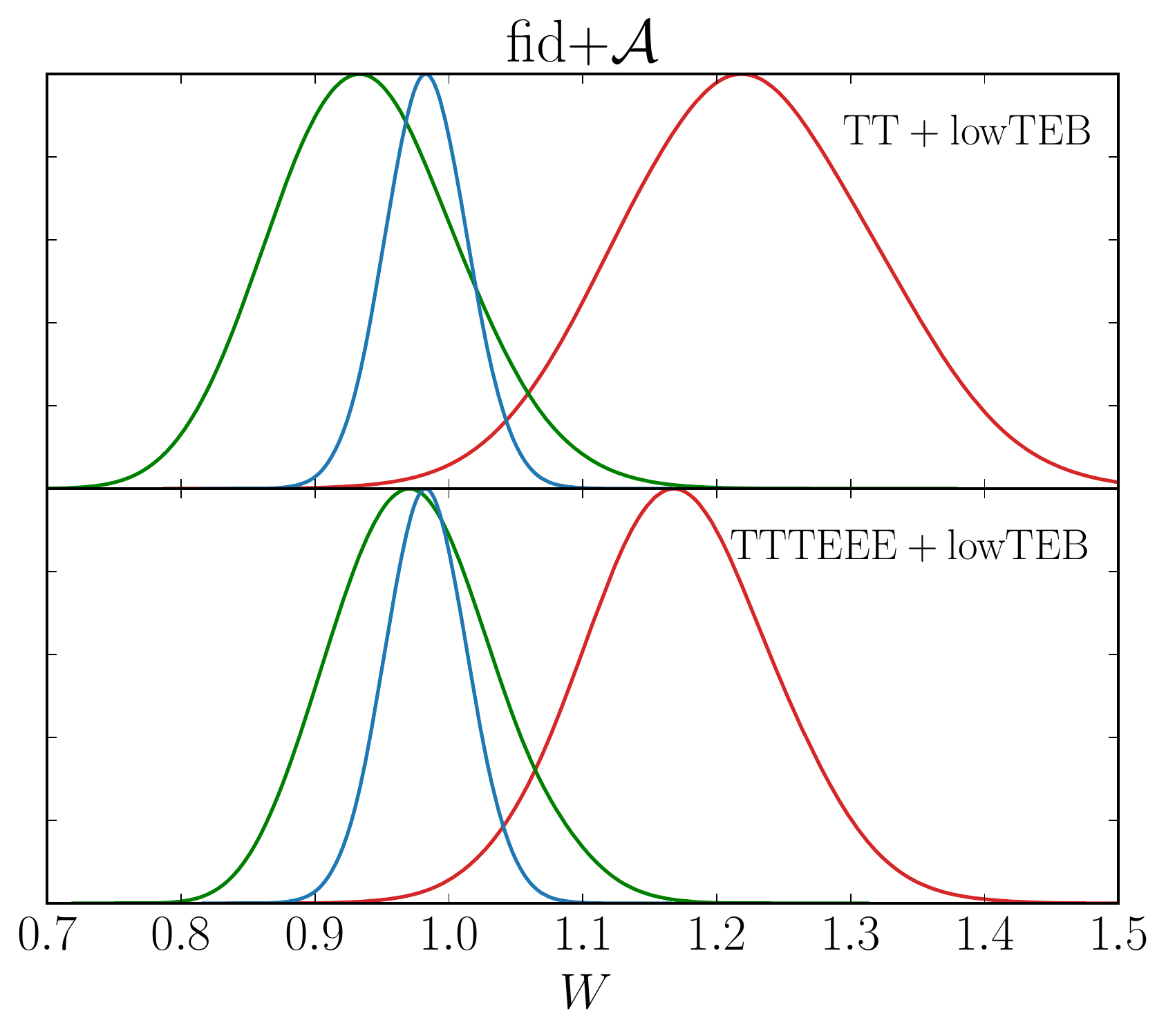}
\cprotect\caption{Posterior probability distributions for the lensing tension parameter $W$
as in Fig.~\ref{fig:tension_gaussianity} but allowing only for amplitude changes
with  the {fid+$\mathcal{A}$} model.  
}
\label{fig:W_amplitude}
\end{figure}

We therefore model the lensing potential as
\be
	C_L^\PP = \mathcal{A} C^\PP_{L,\mathrm{fid}} ,
\ee
below we refer to this model as ``fid + $\mathcal{A}$'' but recall that the fiducial model
is set by the best fit $\Lambda$CDM parameters in Tab.~\ref{tab:fiducial}.
Note that this is different from the standard $A_L$ and $A_\PP$\cprotect\footnote{$A_\PP$ is a
parameter that scales the lensing potential used in the \verb|PP| but not the one used in
the \verb|TT| likelihood.}  parameters in that the
amplitude multiplies a fixed fiducial model.
Constraints on $W$ from these two data sets are shown in
Fig.~\ref{fig:W_amplitude} (top).
Comparing these two constraints leads to a tension of 2.4$\sigma$, the same as the
model-independent tension derived in the previous section. When adding polarization data,
the tension evaluates to the same 2.4$\sigma$, slightly more than the model-independent
value; see Fig.~\ref{fig:W_amplitude} for changes in the posteriors.
Comparing instead constraints on $\mathcal{A}$ directly leads to the same tension
significance both with and without polarization. The good agreement with constraints on
$\mathcal{A}$ gives further evidence that $W$ is a powerful and robust tension indicator,
even though it is constructed from
a single PC.

With $W$ we can also compare the predictions from  the unlensed parameters
$\tilde \theta_A$ in the same {fid+$\mathcal{A}$} context.   
These correspond to the green curves in
Fig.~\ref{fig:W_amplitude} and when compared with their red counterparts evaluate to
internal tensions significant at 2.4$\sigma$ for \verb|TT+lowTEB|, respectively
2.1$\sigma$ for \verb|TTTEEE+lowTEB|.

\begin{figure}
\center
\includegraphics[width = 0.49 \textwidth]{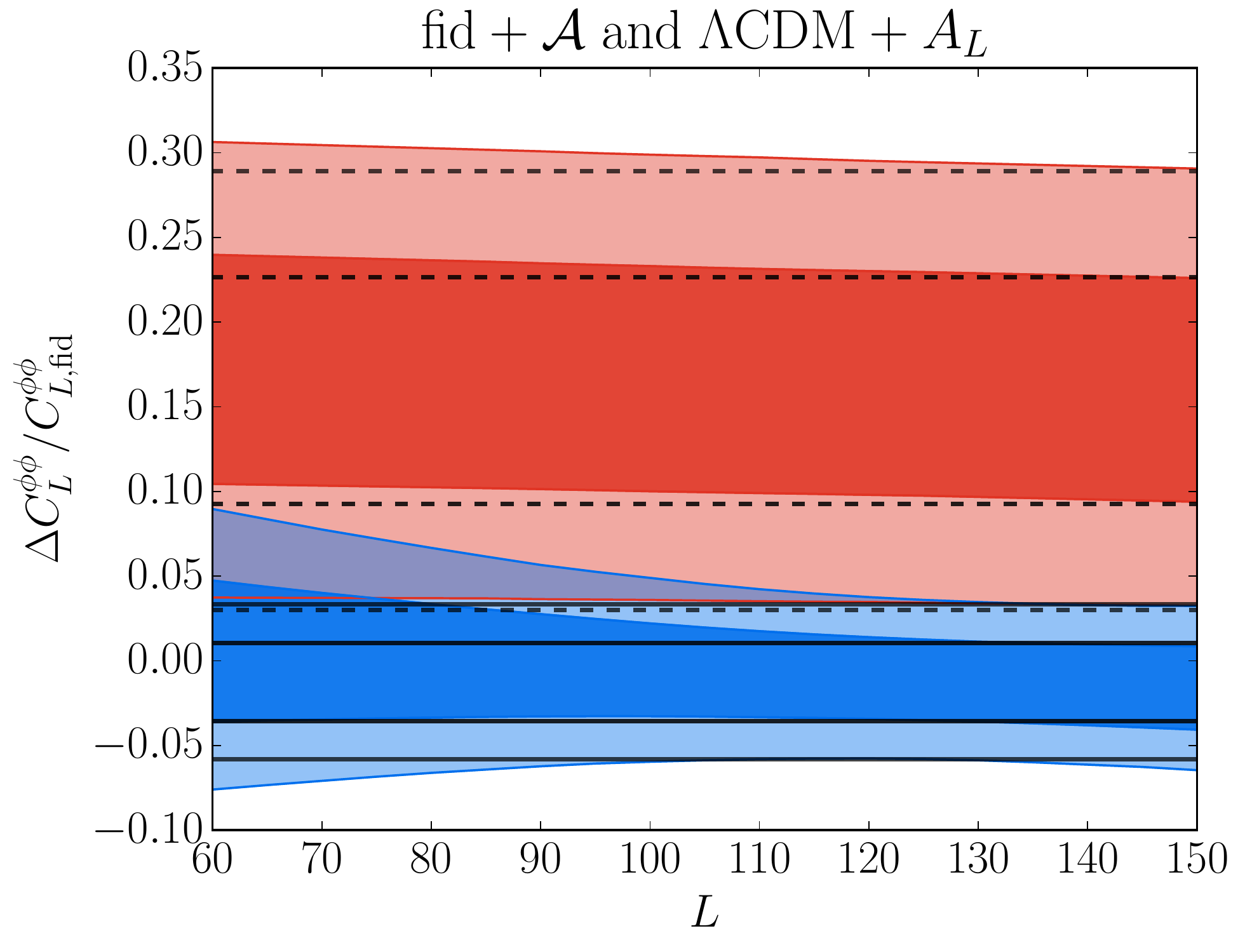}
\cprotect\caption{Constraints on  $\Delta C_{L}^\PP/ C_{L, \rm fid}^\PP$
 as in Fig.~\ref{fig:cpp_posterior}, but for amplitude and
$\Lambda$CDM shape variations.  The black lines show
results from \verb|TT+lowTEB| (dashed) or \verb|PP| (solid) within the
fid+$\mathcal{A}$ for amplitude variations. 
Filled contours are determined from \verb|TT+lowT| 
within $\Lambda$CDM+$A_L$ with a $\tau$ prior (red) and separately from \verb|PP|  with
$\Lambda$CDM freedom on the amplitude and shape but with fixed $\tau, \theta_*$ and a
prior on $\Omega_b h^2$ and $n_s$ (blue).  Constraints, especially at low $L$ from the
latter, depend on which $\Lambda$CDM parameters are allowed to separately vary.
}
\label{fig:aghanim_cpp_posterior_calA}
\end{figure}
Within the {fid+$\mathcal{A}$} model considered here, constraints
on any $C_L^\PP$ show the same significance of the tension {as} $\mathcal{A}$ (see Fig.~\ref{fig:aghanim_cpp_posterior_calA}, solid vs.~dashed lines).
Nonetheless, there are subtleties in using $C_L^\PP$ itself as a tension indicator 
beyond $\mathcal{A}$, even for parameterizations that are motivated by $\Lambda$CDM.
As we argued above, to quote tension in $C_L^\PP$ measurements, one
has to exactly specify how much freedom in the lensing potential is allowed; in the PC
case, this turns into a sensitivity to the number of PCs involved.

For variations motivated by $\Lambda$CDM, one has to carefully specify which parameters
are allowed to independently vary between the reconstruction and CMB power spectra
analyses.
For example, let us take the case of comparing the temperature power spectrum
\verb|TT+lowT|
and lens reconstruction measurements as commonly considered in the literature.  For the former case we further 
take the usual $\Lambda$CDM+$A_L$ approach which allows some variation in the
shape of the lensing power spectrum through the cosmological parameters.
Without the low-$\ell$ polarization data,
we must specify the  prior on $\tau$ since it controls $A_s$ through the
measured amplitude of the temperature peaks.   
   For definiteness let us take
a Gaussian prior of $\tau=0.07\pm 0.02$.   

Given  reconstruction data alone, $\Lambda$CDM
allows both amplitude and larger shape changes since the cosmological parameters are not
constrained by  CMB power spectra. For definiteness, let
us take the joint posterior of the $\Lambda$CDM parameters $\ln A_s, n_s,\Omega_c h^2,\Omega_b
h^2$ with Gaussian priors $\Omega_b h^2 = 0.0223 \pm 0.0009$ and $n_s = 0.96 \pm
0.02$.
Constraints on $C_L^\PP$ when allowing these
variations are shown in Fig.~\ref{fig:aghanim_cpp_posterior_calA} (red vs.~blue contours).
Note that as $L$ decreases, these additional shape variations in $C_L^\PP$ weaken the
apparent tension.

When one compares
constraints on $C_{100}^\PP$, this is the technique used in
Ref.~\cite{Aghanim:2016sns}\footnote{Marius Millea, private communication}. 
  At $L=100$ the shape variation only has a mild effect that is
further reduced by the shift in both contours so for $C_{100}^\PP$ we retain a tension significance of
2.4$\sigma$.\footnote{The small difference from the value 2.3$\sigma$ quoted in
Ref.~\cite{Aghanim:2016sns} can be caused by different analysis choices.}  

At lower $L$ the shape variations become more important.  
On the lensing reconstruction side, the $\Lambda$CDM parameters are not
well constrained and allow values that are inconsistent with the unlensed CMB{; this
leads to the shape variations noticeable at low $L$ in Fig.~\ref{fig:aghanim_cpp_posterior_calA}}.
{Had we allowed {even} larger freedom
in $C_L^\PP$ by changing say the prior on $n_s$ on the reconstruction side, the apparent
tension would further degrade. 
On the \verb|TT+lowT| {side the $\Lambda$CDM parameters are very well constrained,
leaving less of an effect on the shape of $C_L^\PP$ and the contours mostly reflect
uncertainty in the amplitude through $A_L$.}
However, when more freedom is granted to the lensing potential, the apparently
strong constraints at low $L$ degrade (see Fig.~\ref{fig:cpp_posterior}), also decreasing
the tension.
It is thus important to carefully specify the model freedom 
in using $C_L^\PP$ as a statistic.  This problem is largely removed by using
$W$ which has the same meaning in all models.

\section{Discussion}
\label{sec:discussion}

In this work we  separate the lensing information from CMB power spectra 
from the cosmological parameters that control the high redshift physics and thereby
illuminate the so-called ``lensing tensions'' in
the Planck CMB data.  
By modeling the principal components of the lens potential $C_L^\PP$ given the Planck CMB power spectra, we isolate the one aspect that is constrained by the data in a model independent
way.  We then compare this constraint to results from lens reconstruction through the 4 principal components that it constraints to test whether  variations in $C_L^\PP$ beyond $\Lambda$CDM can
relax tension between the two different sources of lensing information.

We show that the tension remains between  the temperature and lensing reconstruction
determinations of $C_L^\PP$  even 
beyond $\Lambda$CDM. 
Previous studies of the Planck lensing anomaly \cite{Ade:2015xua,
Aghanim:2016sns,Addison:2015wyg, Ade:2013zuv} considered only the addition of 
 changes in the amplitude of the
gravitational lensing potential from $\Lambda$CDM predictions.  Our technique extends
these studies and clarifies the nature of the tension by extracting direct constraints
on $C_L^\PP$, which obviates the need for directly specifying $\Lambda$CDM parameters
in interpreting tension, 
and allowing shape variations from $\Lambda$CDM.

Even allowing for shape and amplitude variations beyond $\Lambda$CDM, the
tension between temperature and lensing reconstruction remains at a level of 
2.4$\sigma$, essentially the same as with amplitude variations alone.  
The significance decreases mildly to 
2.2$\sigma$ when polarization data are taken into account unlike in the case of amplitude
variations;  this drop is driven by
preference of the TE data for less lensing.

We evaluate these tension significances x by using a simple
{difference of the means} statistic {tension} on a {simple function of the first
principal component}.
For the Planck 2015 data which measure only a single aspect
of lensing from CMB power spectra, this provides a simple but powerful, robust, and
lensing model independent quantification of tension.   
Our technique can be easily applied to future CMB data
sets, where more and mutually correlated aspects are measured \cite{Motloch:2016zsl, Motloch:2017rlk}, with a suitable generalization of
tension statistics (e.g. \cite{Kullback:1951aaa}).

This tension is driven by the multipole range  $\ell \sim 1250-1500$ in the TT data 
which prefers smoother acoustic peaks than predicted by the standard physics
at recombination and lensing reconstruction.  While new physics at recombination
could in principle relieve tension, it cannot be relieved by adjusting the relativistic
degrees of freedom through  $N_\mathrm{eff}$.  

By separating information in the lensed CMB power spectra into  lensing and unlensed components, we also enable a consistency check on
the  $\Lambda$CDM cosmological model.  Because the $\Lambda$CDM
prediction based on constraints to the unlensed CMB is consistent with the lensing
reconstruction constraints, the internal consistency check fails at similar significance
as the comparison of the temperature -- lensing reconstruction determinations of lensing
potential. Addition of polarization data again decreases the significance of the tension,
more so this time due to preference for high $\Omega_c h^2$ in the TE data.

While these tensions may point to systematic errors or a statistical fluke that is resolved
by more data and improved data reduction, our technique of extracting direct constraints on the lensing
potential from CMB power spectra data should continue to provide a robust and powerful tool 
for testing the consistency of $\Lambda$CDM and searching for new physics in the future.

\acknowledgements{
We thank Niayesh Afshordi, Neal Dalal, Marius Millea and Marco Raveri for useful discussions.
This work was
supported by NASA ATP NNX15AK22G and the Kavli
Institute for Cosmological Physics at the University of Chicago through grant NSF
PHY-1125897 and an endowment from the Kavli Foundation and its founder Fred Kavli.  WH was
additionally supported by   U.S.~Dept.\ of Energy contract DE-FG02-13ER41958  and the Simons Foundation.  
We acknowledge use of the CAMB and CosmoMC software packages. 
This work was completed in part with
resources provided by the University of Chicago Research Computing Center. 
PM thanks the Perimeter Institute for Theoretical Physics where part of this work was
performed. Research at Perimeter Institute is supported by the Government of Canada
through the Department of Innovation, Science and Economic Development and by the Province
of Ontario through the Ministry of Research and Innovation.  
}
\bibliography{plancklens}

\begin{thebibliography}{35}%
\makeatletter
\providecommand \@ifxundefined [1]{%
 \@ifx{#1\undefined}
}%
\providecommand \@ifnum [1]{%
 \ifnum #1\expandafter \@firstoftwo
 \else \expandafter \@secondoftwo
 \fi
}%
\providecommand \@ifx [1]{%
 \ifx #1\expandafter \@firstoftwo
 \else \expandafter \@secondoftwo
 \fi
}%
\providecommand \natexlab [1]{#1}%
\providecommand \enquote  [1]{``#1''}%
\providecommand \bibnamefont  [1]{#1}%
\providecommand \bibfnamefont [1]{#1}%
\providecommand \citenamefont [1]{#1}%
\providecommand \href@noop [0]{\@secondoftwo}%
\providecommand \href [0]{\begingroup \@sanitize@url \@href}%
\providecommand \@href[1]{\@@startlink{#1}\@@href}%
\providecommand \@@href[1]{\endgroup#1\@@endlink}%
\providecommand \@sanitize@url [0]{\catcode `\\12\catcode `\$12\catcode
  `\&12\catcode `\#12\catcode `\^12\catcode `\_12\catcode `\%12\relax}%
\providecommand \@@startlink[1]{}%
\providecommand \@@endlink[0]{}%
\providecommand \url  [0]{\begingroup\@sanitize@url \@url }%
\providecommand \@url [1]{\endgroup\@href {#1}{\urlprefix }}%
\providecommand \urlprefix  [0]{URL }%
\providecommand \Eprint [0]{\href }%
\providecommand \doibase [0]{http://dx.doi.org/}%
\providecommand \selectlanguage [0]{\@gobble}%
\providecommand \bibinfo  [0]{\@secondoftwo}%
\providecommand \bibfield  [0]{\@secondoftwo}%
\providecommand \translation [1]{[#1]}%
\providecommand \BibitemOpen [0]{}%
\providecommand \bibitemStop [0]{}%
\providecommand \bibitemNoStop [0]{.\EOS\space}%
\providecommand \EOS [0]{\spacefactor3000\relax}%
\providecommand \BibitemShut  [1]{\csname bibitem#1\endcsname}%
\let\auto@bib@innerbib\@empty
\bibitem [{\citenamefont {Lewis}\ and\ \citenamefont
  {Challinor}(2006)}]{Lewis:2006fu}%
  \BibitemOpen
  \bibfield  {author} {\bibinfo {author} {\bibfnamefont {A.}~\bibnamefont
  {Lewis}}\ and\ \bibinfo {author} {\bibfnamefont {A.}~\bibnamefont
  {Challinor}},\ }\href {\doibase 10.1016/j.physrep.2006.03.002} {\bibfield
  {journal} {\bibinfo  {journal} {Phys. Rept.}\ }\textbf {\bibinfo {volume}
  {429}},\ \bibinfo {pages} {1} (\bibinfo {year} {2006})},\ \Eprint
  {http://arxiv.org/abs/astro-ph/0601594} {arXiv:astro-ph/0601594 [astro-ph]}
  \BibitemShut {NoStop}%
\bibitem [{\citenamefont {Smith}\ \emph {et~al.}(2007)\citenamefont {Smith},
  \citenamefont {Zahn},\ and\ \citenamefont {Dore}}]{Smith:2007rg}%
  \BibitemOpen
  \bibfield  {author} {\bibinfo {author} {\bibfnamefont {K.~M.}\ \bibnamefont
  {Smith}}, \bibinfo {author} {\bibfnamefont {O.}~\bibnamefont {Zahn}}, \ and\
  \bibinfo {author} {\bibfnamefont {O.}~\bibnamefont {Dore}},\ }\href {\doibase
  10.1103/PhysRevD.76.043510} {\bibfield  {journal} {\bibinfo  {journal} {Phys.
  Rev.}\ }\textbf {\bibinfo {volume} {D76}},\ \bibinfo {pages} {043510}
  (\bibinfo {year} {2007})},\ \Eprint {http://arxiv.org/abs/0705.3980}
  {arXiv:0705.3980 [astro-ph]} \BibitemShut {NoStop}%
\bibitem [{\citenamefont {Hirata}\ \emph {et~al.}(2008)\citenamefont {Hirata},
  \citenamefont {Ho}, \citenamefont {Padmanabhan}, \citenamefont {Seljak},\
  and\ \citenamefont {Bahcall}}]{Hirata08}%
  \BibitemOpen
  \bibfield  {author} {\bibinfo {author} {\bibfnamefont {C.~M.}\ \bibnamefont
  {Hirata}}, \bibinfo {author} {\bibfnamefont {S.}~\bibnamefont {Ho}}, \bibinfo
  {author} {\bibfnamefont {N.}~\bibnamefont {Padmanabhan}}, \bibinfo {author}
  {\bibfnamefont {U.}~\bibnamefont {Seljak}}, \ and\ \bibinfo {author}
  {\bibfnamefont {N.~A.}\ \bibnamefont {Bahcall}},\ }\href {\doibase
  10.1103/PhysRevD.78.043520} {\bibfield  {journal} {\bibinfo  {journal} {Phys.
  Rev. D}\ }\textbf {\bibinfo {volume} {78}},\ \bibinfo {pages} {043520}
  (\bibinfo {year} {2008})}\BibitemShut {NoStop}%
\bibitem [{\citenamefont {Hanson}\ \emph {et~al.}(2013)\citenamefont {Hanson}
  \emph {et~al.}}]{Hanson:2013hsb}%
  \BibitemOpen
  \bibfield  {author} {\bibinfo {author} {\bibfnamefont {D.}~\bibnamefont
  {Hanson}} \emph {et~al.} (\bibinfo {collaboration} {SPTpol}),\ }\href
  {\doibase 10.1103/PhysRevLett.111.141301} {\bibfield  {journal} {\bibinfo
  {journal} {Phys. Rev. Lett.}\ }\textbf {\bibinfo {volume} {111}},\ \bibinfo
  {pages} {141301} (\bibinfo {year} {2013})},\ \Eprint
  {http://arxiv.org/abs/1307.5830} {arXiv:1307.5830 [astro-ph.CO]} \BibitemShut
  {NoStop}%
\bibitem [{\citenamefont {Das}\ \emph {et~al.}(2011)\citenamefont {Das} \emph
  {et~al.}}]{Das:2011ak}%
  \BibitemOpen
  \bibfield  {author} {\bibinfo {author} {\bibfnamefont {S.}~\bibnamefont
  {Das}} \emph {et~al.},\ }\href {\doibase 10.1103/PhysRevLett.107.021301}
  {\bibfield  {journal} {\bibinfo  {journal} {Phys. Rev. Lett.}\ }\textbf
  {\bibinfo {volume} {107}},\ \bibinfo {pages} {021301} (\bibinfo {year}
  {2011})},\ \Eprint {http://arxiv.org/abs/1103.2124} {arXiv:1103.2124
  [astro-ph.CO]} \BibitemShut {NoStop}%
\bibitem [{\citenamefont {Keisler}\ \emph {et~al.}(2011)\citenamefont {Keisler}
  \emph {et~al.}}]{Keisler:2011aw}%
  \BibitemOpen
  \bibfield  {author} {\bibinfo {author} {\bibfnamefont {R.}~\bibnamefont
  {Keisler}} \emph {et~al.},\ }\href {\doibase 10.1088/0004-637X/743/1/28}
  {\bibfield  {journal} {\bibinfo  {journal} {Astrophys. J.}\ }\textbf
  {\bibinfo {volume} {743}},\ \bibinfo {pages} {28} (\bibinfo {year} {2011})},\
  \Eprint {http://arxiv.org/abs/1105.3182} {arXiv:1105.3182 [astro-ph.CO]}
  \BibitemShut {NoStop}%
\bibitem [{\citenamefont {{Ade}}\ \emph {et~al.}(2014)\citenamefont {{Ade}}
  \emph {et~al.}}]{Planck2013XVII}%
  \BibitemOpen
  \bibfield  {author} {\bibinfo {author} {\bibfnamefont {P.~A.~R.}\
  \bibnamefont {{Ade}}} \emph {et~al.} (\bibinfo {collaboration} {The Planck
  Collaboration}),\ }\href {\doibase 10.1051/0004-6361/201321543} {\bibfield
  {journal} {\bibinfo  {journal} {Astron. \& Astrophys.}\ }\textbf {\bibinfo
  {volume} {571}},\ \bibinfo {eid} {A17} (\bibinfo {year} {2014})},\ \Eprint
  {http://arxiv.org/abs/1303.5077} {arXiv:1303.5077} \BibitemShut {NoStop}%
\bibitem [{\citenamefont {Keisler}\ \emph {et~al.}(2015)\citenamefont {Keisler}
  \emph {et~al.}}]{Keisler:2015hfa}%
  \BibitemOpen
  \bibfield  {author} {\bibinfo {author} {\bibfnamefont {R.}~\bibnamefont
  {Keisler}} \emph {et~al.} (\bibinfo {collaboration} {SPT}),\ }\href {\doibase
  10.1088/0004-637X/807/2/151} {\bibfield  {journal} {\bibinfo  {journal}
  {Astrophys. J.}\ }\textbf {\bibinfo {volume} {807}},\ \bibinfo {pages} {151}
  (\bibinfo {year} {2015})},\ \Eprint {http://arxiv.org/abs/1503.02315}
  {arXiv:1503.02315 [astro-ph.CO]} \BibitemShut {NoStop}%
\bibitem [{\citenamefont {Ade}\ \emph {et~al.}(2016{\natexlab{a}})\citenamefont
  {Ade} \emph {et~al.}}]{Ade:2015zua}%
  \BibitemOpen
  \bibfield  {author} {\bibinfo {author} {\bibfnamefont {P.~A.~R.}\
  \bibnamefont {Ade}} \emph {et~al.} (\bibinfo {collaboration} {Planck}),\
  }\href {\doibase 10.1051/0004-6361/201525941} {\bibfield  {journal} {\bibinfo
   {journal} {Astron. Astrophys.}\ }\textbf {\bibinfo {volume} {594}},\
  \bibinfo {pages} {A15} (\bibinfo {year} {2016}{\natexlab{a}})},\ \Eprint
  {http://arxiv.org/abs/1502.01591} {arXiv:1502.01591 [astro-ph.CO]}
  \BibitemShut {NoStop}%
\bibitem [{\citenamefont {Ade}\ \emph {et~al.}(2016{\natexlab{b}})\citenamefont
  {Ade} \emph {et~al.}}]{Array:2016afx}%
  \BibitemOpen
  \bibfield  {author} {\bibinfo {author} {\bibfnamefont {P.~A.~R.}\
  \bibnamefont {Ade}} \emph {et~al.} (\bibinfo {collaboration} {BICEP2, Keck
  Array}),\ }\href@noop {} {\  (\bibinfo {year} {2016}{\natexlab{b}})},\
  \Eprint {http://arxiv.org/abs/1606.01968} {arXiv:1606.01968 [astro-ph.CO]}
  \BibitemShut {NoStop}%
\bibitem [{\citenamefont {{Sherwin}}\ \emph {et~al.}(2016)\citenamefont
  {{Sherwin}} \emph {et~al.}}]{Sherwin2016}%
  \BibitemOpen
  \bibfield  {author} {\bibinfo {author} {\bibfnamefont {B.~D.}\ \bibnamefont
  {{Sherwin}}} \emph {et~al.},\ }\href@noop {} {\bibfield  {journal} {\bibinfo
  {journal} {ArXiv e-prints}\ } (\bibinfo {year} {2016})},\ \Eprint
  {http://arxiv.org/abs/1611.09753} {arXiv:1611.09753} \BibitemShut {NoStop}%
\bibitem [{\citenamefont {Seljak}(1996)}]{Seljak:1995ve}%
  \BibitemOpen
  \bibfield  {author} {\bibinfo {author} {\bibfnamefont {U.}~\bibnamefont
  {Seljak}},\ }\href {\doibase 10.1086/177218} {\bibfield  {journal} {\bibinfo
  {journal} {Astrophys. J.}\ }\textbf {\bibinfo {volume} {463}},\ \bibinfo
  {pages} {1} (\bibinfo {year} {1996})},\ \Eprint
  {http://arxiv.org/abs/astro-ph/9505109} {arXiv:astro-ph/9505109 [astro-ph]}
  \BibitemShut {NoStop}%
\bibitem [{\citenamefont {Zaldarriaga}(2000)}]{Zaldarriaga:2000ud}%
  \BibitemOpen
  \bibfield  {author} {\bibinfo {author} {\bibfnamefont {M.}~\bibnamefont
  {Zaldarriaga}},\ }\href {\doibase 10.1103/PhysRevD.62.063510} {\bibfield
  {journal} {\bibinfo  {journal} {Phys. Rev.}\ }\textbf {\bibinfo {volume}
  {D62}},\ \bibinfo {pages} {063510} (\bibinfo {year} {2000})},\ \Eprint
  {http://arxiv.org/abs/astro-ph/9910498} {arXiv:astro-ph/9910498 [astro-ph]}
  \BibitemShut {NoStop}%
\bibitem [{\citenamefont {Hu}(2001)}]{Hu:2001fa}%
  \BibitemOpen
  \bibfield  {author} {\bibinfo {author} {\bibfnamefont {W.}~\bibnamefont
  {Hu}},\ }\href {\doibase 10.1103/PhysRevD.64.083005} {\bibfield  {journal}
  {\bibinfo  {journal} {Phys. Rev.}\ }\textbf {\bibinfo {volume} {D64}},\
  \bibinfo {pages} {083005} (\bibinfo {year} {2001})},\ \Eprint
  {http://arxiv.org/abs/astro-ph/0105117} {arXiv:astro-ph/0105117 [astro-ph]}
  \BibitemShut {NoStop}%
\bibitem [{\citenamefont {Hu}\ and\ \citenamefont {Okamoto}(2002)}]{Hu:2001kj}%
  \BibitemOpen
  \bibfield  {author} {\bibinfo {author} {\bibfnamefont {W.}~\bibnamefont
  {Hu}}\ and\ \bibinfo {author} {\bibfnamefont {T.}~\bibnamefont {Okamoto}},\
  }\href {\doibase 10.1086/341110} {\bibfield  {journal} {\bibinfo  {journal}
  {Astrophys. J.}\ }\textbf {\bibinfo {volume} {574}},\ \bibinfo {pages} {566}
  (\bibinfo {year} {2002})},\ \Eprint {http://arxiv.org/abs/astro-ph/0111606}
  {arXiv:astro-ph/0111606 [astro-ph]} \BibitemShut {NoStop}%
\bibitem [{\citenamefont {Okamoto}\ and\ \citenamefont
  {Hu}(2003)}]{Okamoto:2003zw}%
  \BibitemOpen
  \bibfield  {author} {\bibinfo {author} {\bibfnamefont {T.}~\bibnamefont
  {Okamoto}}\ and\ \bibinfo {author} {\bibfnamefont {W.}~\bibnamefont {Hu}},\
  }\href {\doibase 10.1103/PhysRevD.67.083002} {\bibfield  {journal} {\bibinfo
  {journal} {Phys. Rev.}\ }\textbf {\bibinfo {volume} {D67}},\ \bibinfo {pages}
  {083002} (\bibinfo {year} {2003})},\ \Eprint
  {http://arxiv.org/abs/astro-ph/0301031} {arXiv:astro-ph/0301031 [astro-ph]}
  \BibitemShut {NoStop}%
\bibitem [{\citenamefont {Hirata}\ and\ \citenamefont
  {Seljak}(2003)}]{Hirata:2003ka}%
  \BibitemOpen
  \bibfield  {author} {\bibinfo {author} {\bibfnamefont {C.~M.}\ \bibnamefont
  {Hirata}}\ and\ \bibinfo {author} {\bibfnamefont {U.}~\bibnamefont
  {Seljak}},\ }\href {\doibase 10.1103/PhysRevD.68.083002} {\bibfield
  {journal} {\bibinfo  {journal} {Phys. Rev.}\ }\textbf {\bibinfo {volume}
  {D68}},\ \bibinfo {pages} {083002} (\bibinfo {year} {2003})},\ \Eprint
  {http://arxiv.org/abs/astro-ph/0306354} {arXiv:astro-ph/0306354 [astro-ph]}
  \BibitemShut {NoStop}%
\bibitem [{\citenamefont {Smith}\ \emph {et~al.}(2012)\citenamefont {Smith},
  \citenamefont {Hanson}, \citenamefont {LoVerde}, \citenamefont {Hirata},\
  and\ \citenamefont {Zahn}}]{Smith:2010gu}%
  \BibitemOpen
  \bibfield  {author} {\bibinfo {author} {\bibfnamefont {K.~M.}\ \bibnamefont
  {Smith}}, \bibinfo {author} {\bibfnamefont {D.}~\bibnamefont {Hanson}},
  \bibinfo {author} {\bibfnamefont {M.}~\bibnamefont {LoVerde}}, \bibinfo
  {author} {\bibfnamefont {C.~M.}\ \bibnamefont {Hirata}}, \ and\ \bibinfo
  {author} {\bibfnamefont {O.}~\bibnamefont {Zahn}},\ }\href {\doibase
  10.1088/1475-7516/2012/06/014} {\bibfield  {journal} {\bibinfo  {journal}
  {JCAP}\ }\textbf {\bibinfo {volume} {1206}},\ \bibinfo {pages} {014}
  (\bibinfo {year} {2012})},\ \Eprint {http://arxiv.org/abs/1010.0048}
  {arXiv:1010.0048 [astro-ph.CO]} \BibitemShut {NoStop}%
\bibitem [{\citenamefont {Ade}\ \emph {et~al.}(2016{\natexlab{c}})\citenamefont
  {Ade} \emph {et~al.}}]{Ade:2015xua}%
  \BibitemOpen
  \bibfield  {author} {\bibinfo {author} {\bibfnamefont {P.~A.~R.}\
  \bibnamefont {Ade}} \emph {et~al.} (\bibinfo {collaboration} {Planck}),\
  }\href {\doibase 10.1051/0004-6361/201525830} {\bibfield  {journal} {\bibinfo
   {journal} {Astron. Astrophys.}\ }\textbf {\bibinfo {volume} {594}},\
  \bibinfo {pages} {A13} (\bibinfo {year} {2016}{\natexlab{c}})},\ \Eprint
  {http://arxiv.org/abs/1502.01589} {arXiv:1502.01589 [astro-ph.CO]}
  \BibitemShut {NoStop}%
\bibitem [{\citenamefont {Aghanim}\ \emph {et~al.}(2017)\citenamefont {Aghanim}
  \emph {et~al.}}]{Aghanim:2016sns}%
  \BibitemOpen
  \bibfield  {author} {\bibinfo {author} {\bibfnamefont {N.}~\bibnamefont
  {Aghanim}} \emph {et~al.} (\bibinfo {collaboration} {Planck}),\ }\href
  {\doibase 10.1051/0004-6361/201629504} {\bibfield  {journal} {\bibinfo
  {journal} {Astron. Astrophys.}\ }\textbf {\bibinfo {volume} {607}},\ \bibinfo
  {pages} {A95} (\bibinfo {year} {2017})},\ \Eprint
  {http://arxiv.org/abs/1608.02487} {arXiv:1608.02487 [astro-ph.CO]}
  \BibitemShut {NoStop}%
\bibitem [{\citenamefont {Addison}\ \emph {et~al.}(2016)\citenamefont
  {Addison}, \citenamefont {Huang}, \citenamefont {Watts}, \citenamefont
  {Bennett}, \citenamefont {Halpern}, \citenamefont {Hinshaw},\ and\
  \citenamefont {Weiland}}]{Addison:2015wyg}%
  \BibitemOpen
  \bibfield  {author} {\bibinfo {author} {\bibfnamefont {G.~E.}\ \bibnamefont
  {Addison}}, \bibinfo {author} {\bibfnamefont {Y.}~\bibnamefont {Huang}},
  \bibinfo {author} {\bibfnamefont {D.~J.}\ \bibnamefont {Watts}}, \bibinfo
  {author} {\bibfnamefont {C.~L.}\ \bibnamefont {Bennett}}, \bibinfo {author}
  {\bibfnamefont {M.}~\bibnamefont {Halpern}}, \bibinfo {author} {\bibfnamefont
  {G.}~\bibnamefont {Hinshaw}}, \ and\ \bibinfo {author} {\bibfnamefont
  {J.~L.}\ \bibnamefont {Weiland}},\ }\href {\doibase
  10.3847/0004-637X/818/2/132} {\bibfield  {journal} {\bibinfo  {journal}
  {Astrophys. J.}\ }\textbf {\bibinfo {volume} {818}},\ \bibinfo {pages} {132}
  (\bibinfo {year} {2016})},\ \Eprint {http://arxiv.org/abs/1511.00055}
  {arXiv:1511.00055 [astro-ph.CO]} \BibitemShut {NoStop}%
\bibitem [{\citenamefont {Ade}\ \emph {et~al.}(2014)\citenamefont {Ade} \emph
  {et~al.}}]{Ade:2013zuv}%
  \BibitemOpen
  \bibfield  {author} {\bibinfo {author} {\bibfnamefont {P.~A.~R.}\
  \bibnamefont {Ade}} \emph {et~al.} (\bibinfo {collaboration} {Planck}),\
  }\href {\doibase 10.1051/0004-6361/201321591} {\bibfield  {journal} {\bibinfo
   {journal} {Astron. Astrophys.}\ }\textbf {\bibinfo {volume} {571}},\
  \bibinfo {pages} {A16} (\bibinfo {year} {2014})},\ \Eprint
  {http://arxiv.org/abs/1303.5076} {arXiv:1303.5076 [astro-ph.CO]} \BibitemShut
  {NoStop}%
\bibitem [{\citenamefont {Carron}\ \emph {et~al.}(2017)\citenamefont {Carron},
  \citenamefont {Lewis},\ and\ \citenamefont {Challinor}}]{Carron:2017vfg}%
  \BibitemOpen
  \bibfield  {author} {\bibinfo {author} {\bibfnamefont {J.}~\bibnamefont
  {Carron}}, \bibinfo {author} {\bibfnamefont {A.}~\bibnamefont {Lewis}}, \
  and\ \bibinfo {author} {\bibfnamefont {A.}~\bibnamefont {Challinor}},\ }\href
  {\doibase 10.1088/1475-7516/2017/05/035} {\bibfield  {journal} {\bibinfo
  {journal} {JCAP}\ }\textbf {\bibinfo {volume} {1705}},\ \bibinfo {pages}
  {035} (\bibinfo {year} {2017})},\ \Eprint {http://arxiv.org/abs/1701.01712}
  {arXiv:1701.01712 [astro-ph.CO]} \BibitemShut {NoStop}%
\bibitem [{\citenamefont {Motloch}\ \emph {et~al.}(2017)\citenamefont
  {Motloch}, \citenamefont {Hu},\ and\ \citenamefont
  {Benoit-L\'{e}vy}}]{Motloch:2016zsl}%
  \BibitemOpen
  \bibfield  {author} {\bibinfo {author} {\bibfnamefont {P.}~\bibnamefont
  {Motloch}}, \bibinfo {author} {\bibfnamefont {W.}~\bibnamefont {Hu}}, \ and\
  \bibinfo {author} {\bibfnamefont {A.}~\bibnamefont {Benoit-L\'{e}vy}},\
  }\href {\doibase 10.1103/PhysRevD.95.043518} {\bibfield  {journal} {\bibinfo
  {journal} {Phys. Rev.}\ }\textbf {\bibinfo {volume} {D95}},\ \bibinfo {pages}
  {043518} (\bibinfo {year} {2017})},\ \Eprint
  {http://arxiv.org/abs/1612.05637} {arXiv:1612.05637 [astro-ph.CO]}
  \BibitemShut {NoStop}%
\bibitem [{\citenamefont {Motloch}\ and\ \citenamefont
  {Hu}(2017)}]{Motloch:2017rlk}%
  \BibitemOpen
  \bibfield  {author} {\bibinfo {author} {\bibfnamefont {P.}~\bibnamefont
  {Motloch}}\ and\ \bibinfo {author} {\bibfnamefont {W.}~\bibnamefont {Hu}},\
  }\href@noop {} {\  (\bibinfo {year} {2017})},\ \Eprint
  {http://arxiv.org/abs/1709.03599} {arXiv:1709.03599 [astro-ph.CO]}
  \BibitemShut {NoStop}%
\bibitem [{\citenamefont {Lewis}\ and\ \citenamefont
  {Bridle}(2002)}]{Lewis:2002ah}%
  \BibitemOpen
  \bibfield  {author} {\bibinfo {author} {\bibfnamefont {A.}~\bibnamefont
  {Lewis}}\ and\ \bibinfo {author} {\bibfnamefont {S.}~\bibnamefont {Bridle}},\
  }\href {\doibase 10.1103/PhysRevD.66.103511} {\bibfield  {journal} {\bibinfo
  {journal} {Phys. Rev.}\ }\textbf {\bibinfo {volume} {D66}},\ \bibinfo {pages}
  {103511} (\bibinfo {year} {2002})},\ \Eprint
  {http://arxiv.org/abs/astro-ph/0205436} {arXiv:astro-ph/0205436 [astro-ph]}
  \BibitemShut {NoStop}%
\bibitem [{\citenamefont {Gelman}\ and\ \citenamefont
  {Rubin}(1992)}]{Gelman:1992zz}%
  \BibitemOpen
  \bibfield  {author} {\bibinfo {author} {\bibfnamefont {A.}~\bibnamefont
  {Gelman}}\ and\ \bibinfo {author} {\bibfnamefont {D.~B.}\ \bibnamefont
  {Rubin}},\ }\href {\doibase 10.1214/ss/1177011136} {\bibfield  {journal}
  {\bibinfo  {journal} {Statist. Sci.}\ }\textbf {\bibinfo {volume} {7}},\
  \bibinfo {pages} {457} (\bibinfo {year} {1992})}\BibitemShut {NoStop}%
\bibitem [{\citenamefont {Smith}\ \emph {et~al.}(2006)\citenamefont {Smith},
  \citenamefont {Hu},\ and\ \citenamefont {Kaplinghat}}]{Smith:2006nk}%
  \BibitemOpen
  \bibfield  {author} {\bibinfo {author} {\bibfnamefont {K.~M.}\ \bibnamefont
  {Smith}}, \bibinfo {author} {\bibfnamefont {W.}~\bibnamefont {Hu}}, \ and\
  \bibinfo {author} {\bibfnamefont {M.}~\bibnamefont {Kaplinghat}},\ }\href
  {\doibase 10.1103/PhysRevD.74.123002} {\bibfield  {journal} {\bibinfo
  {journal} {Phys. Rev.}\ }\textbf {\bibinfo {volume} {D74}},\ \bibinfo {pages}
  {123002} (\bibinfo {year} {2006})},\ \Eprint
  {http://arxiv.org/abs/astro-ph/0607315} {arXiv:astro-ph/0607315 [astro-ph]}
  \BibitemShut {NoStop}%
\bibitem [{\citenamefont {Adam}\ \emph {et~al.}(2016)\citenamefont {Adam} \emph
  {et~al.}}]{Adam:2016hgk}%
  \BibitemOpen
  \bibfield  {author} {\bibinfo {author} {\bibfnamefont {R.}~\bibnamefont
  {Adam}} \emph {et~al.} (\bibinfo {collaboration} {Planck}),\ }\href {\doibase
  10.1051/0004-6361/201628897} {\bibfield  {journal} {\bibinfo  {journal}
  {Astron. Astrophys.}\ }\textbf {\bibinfo {volume} {596}},\ \bibinfo {pages}
  {A108} (\bibinfo {year} {2016})},\ \Eprint {http://arxiv.org/abs/1605.03507}
  {arXiv:1605.03507 [astro-ph.CO]} \BibitemShut {NoStop}%
\bibitem [{\citenamefont {Benoit-Levy}\ \emph {et~al.}(2012)\citenamefont
  {Benoit-Levy}, \citenamefont {Smith},\ and\ \citenamefont
  {Hu}}]{BenoitLevy:2012va}%
  \BibitemOpen
  \bibfield  {author} {\bibinfo {author} {\bibfnamefont {A.}~\bibnamefont
  {Benoit-Levy}}, \bibinfo {author} {\bibfnamefont {K.~M.}\ \bibnamefont
  {Smith}}, \ and\ \bibinfo {author} {\bibfnamefont {W.}~\bibnamefont {Hu}},\
  }\href {\doibase 10.1103/PhysRevD.86.123008} {\bibfield  {journal} {\bibinfo
  {journal} {Phys. Rev.}\ }\textbf {\bibinfo {volume} {D86}},\ \bibinfo {pages}
  {123008} (\bibinfo {year} {2012})},\ \Eprint {http://arxiv.org/abs/1205.0474}
  {arXiv:1205.0474 [astro-ph.CO]} \BibitemShut {NoStop}%
\bibitem [{\citenamefont {Obied}\ \emph {et~al.}(2017)\citenamefont {Obied},
  \citenamefont {Dvorkin}, \citenamefont {Heinrich}, \citenamefont {Hu},\ and\
  \citenamefont {Miranda}}]{Obied:2017tpd}%
  \BibitemOpen
  \bibfield  {author} {\bibinfo {author} {\bibfnamefont {G.}~\bibnamefont
  {Obied}}, \bibinfo {author} {\bibfnamefont {C.}~\bibnamefont {Dvorkin}},
  \bibinfo {author} {\bibfnamefont {C.}~\bibnamefont {Heinrich}}, \bibinfo
  {author} {\bibfnamefont {W.}~\bibnamefont {Hu}}, \ and\ \bibinfo {author}
  {\bibfnamefont {V.}~\bibnamefont {Miranda}},\ }\href {\doibase
  10.1103/PhysRevD.96.083526} {\bibfield  {journal} {\bibinfo  {journal} {Phys.
  Rev.}\ }\textbf {\bibinfo {volume} {D96}},\ \bibinfo {pages} {083526}
  (\bibinfo {year} {2017})},\ \Eprint {http://arxiv.org/abs/1706.09412}
  {arXiv:1706.09412 [astro-ph.CO]} \BibitemShut {NoStop}%
\bibitem [{\citenamefont {{Riess}}\ \emph {et~al.}(2018)\citenamefont
  {{Riess}}, \citenamefont {{Casertano}}, \citenamefont {{Yuan}}, \citenamefont
  {{Macri}}, \citenamefont {{Anderson}}, \citenamefont {{MacKenty}},
  \citenamefont {{Bowers}}, \citenamefont {{Clubb}}, \citenamefont
  {{Filippenko}}, \citenamefont {{Jones}},\ and\ \citenamefont
  {{Tucker}}}]{Riess:2018aaa}%
  \BibitemOpen
  \bibfield  {author} {\bibinfo {author} {\bibfnamefont {A.~G.}\ \bibnamefont
  {{Riess}}}, \bibinfo {author} {\bibfnamefont {S.}~\bibnamefont
  {{Casertano}}}, \bibinfo {author} {\bibfnamefont {W.}~\bibnamefont {{Yuan}}},
  \bibinfo {author} {\bibfnamefont {L.}~\bibnamefont {{Macri}}}, \bibinfo
  {author} {\bibfnamefont {J.}~\bibnamefont {{Anderson}}}, \bibinfo {author}
  {\bibfnamefont {J.~W.}\ \bibnamefont {{MacKenty}}}, \bibinfo {author}
  {\bibfnamefont {J.~B.}\ \bibnamefont {{Bowers}}}, \bibinfo {author}
  {\bibfnamefont {K.~I.}\ \bibnamefont {{Clubb}}}, \bibinfo {author}
  {\bibfnamefont {A.~V.}\ \bibnamefont {{Filippenko}}}, \bibinfo {author}
  {\bibfnamefont {D.~O.}\ \bibnamefont {{Jones}}}, \ and\ \bibinfo {author}
  {\bibfnamefont {B.~E.}\ \bibnamefont {{Tucker}}},\ }\href {\doibase
  10.3847/1538-4357/aaadb7} {\bibfield  {journal} {\bibinfo  {journal} {\apj}\
  }\textbf {\bibinfo {volume} {855}},\ \bibinfo {eid} {136} (\bibinfo {year}
  {2018})},\ \Eprint {http://arxiv.org/abs/1801.01120} {arXiv:1801.01120
  [astro-ph.SR]} \BibitemShut {NoStop}%
\bibitem [{\citenamefont {Hu}\ and\ \citenamefont {Holder}(2003)}]{Hu:2003gh}%
  \BibitemOpen
  \bibfield  {author} {\bibinfo {author} {\bibfnamefont {W.}~\bibnamefont
  {Hu}}\ and\ \bibinfo {author} {\bibfnamefont {G.~P.}\ \bibnamefont
  {Holder}},\ }\href {\doibase 10.1103/PhysRevD.68.023001} {\bibfield
  {journal} {\bibinfo  {journal} {Phys. Rev.}\ }\textbf {\bibinfo {volume}
  {D68}},\ \bibinfo {pages} {023001} (\bibinfo {year} {2003})},\ \Eprint
  {http://arxiv.org/abs/astro-ph/0303400} {arXiv:astro-ph/0303400 [astro-ph]}
  \BibitemShut {NoStop}%
\bibitem [{\citenamefont {Heinrich}\ \emph {et~al.}(2016)\citenamefont
  {Heinrich}, \citenamefont {Miranda},\ and\ \citenamefont
  {Hu}}]{Heinrich:2016ojb}%
  \BibitemOpen
  \bibfield  {author} {\bibinfo {author} {\bibfnamefont {C.~H.}\ \bibnamefont
  {Heinrich}}, \bibinfo {author} {\bibfnamefont {V.}~\bibnamefont {Miranda}}, \
  and\ \bibinfo {author} {\bibfnamefont {W.}~\bibnamefont {Hu}},\ }\href@noop
  {} {\  (\bibinfo {year} {2016})},\ \Eprint {http://arxiv.org/abs/1609.04788}
  {arXiv:1609.04788 [astro-ph.CO]} \BibitemShut {NoStop}%
\bibitem [{\citenamefont {Kullback}\ and\ \citenamefont
  {Leibler}(1951)}]{Kullback:1951aaa}%
  \BibitemOpen
  \bibfield  {author} {\bibinfo {author} {\bibfnamefont {S.}~\bibnamefont
  {Kullback}}\ and\ \bibinfo {author} {\bibfnamefont {R.~A.}\ \bibnamefont
  {Leibler}},\ }\href {\doibase 10.1214/aoms/1177729694} {\bibfield  {journal}
  {\bibinfo  {journal} {Ann. Math. Statist.}\ }\textbf {\bibinfo {volume}
  {22}},\ \bibinfo {pages} {79} (\bibinfo {year} {1951})}\BibitemShut {NoStop}%
\end{thebibliography}%

\end{document}